%% file: ms.tex
\documentclass[12pt,preprint]{aastex}

\def\halpha{\mbox{\rm{H}$\alpha$}}
\def\hbeta{\mbox{\rm{H}$\beta$}}

\def\hdelta{\mbox{\rm{H}$\delta$}}

\begin{document}

\title{Deviations from He~{\sc i} Case~B Recombination Theory
and Extinction Corrections in the Orion Nebula\altaffilmark{1}}

\author{
K. P. M. Blagrave\altaffilmark{2},
P. G. Martin\altaffilmark{2},
R. H. Rubin\altaffilmark{3,4},
R. J. Dufour\altaffilmark{5},
J. A. Baldwin\altaffilmark{6},
J. J. Hester\altaffilmark{7}, and
D. K. Walter\altaffilmark{8}
}

\altaffiltext{1}{Based in part on observations made with the NASA/ESA
{\it Hubble Space Telescope}, obtained at the Space Telescope Science
Institute, which is operated by AURA, Inc., under NASA contract
NAS5-26555}
\altaffiltext{2}{Canadian Institute for Theoretical Astrophysics, University
of Toronto, Toronto, ON, Canada M5S~3H8}
\altaffiltext{3}{NASA/Ames Research Center, Moffett Field, CA 94035}
\altaffiltext{4}{Orion Enterprises}
\altaffiltext{5}{Rice University}
\altaffiltext{6}{Michigan State University}
\altaffiltext{7}{Arizona State University}
\altaffiltext{8}{South Carolina State University}

\begin{abstract}

We are engaged in a comprehensive program to find reliable elemental 
abundances in and to probe the physical structure of the Orion Nebula, the 
brightest and best-resolved H~{\sc ii} region.
In the course of developing a robust extinction correction covering our 
optical and ultraviolet FOS and STIS observations,
we examined the 
decrement within various
series of He~{\sc i} lines. 
The decrements of the $2~^3S-n~^3P$, $2~^3P-n~^3S$ and $3~^3S-n~^3P$
series are not in accord with case~B recombination theory.
None of these anomalous He~{\sc i} decrements
can be explained by extinction, indicating the presence of additional
radiative transfer effects in He~{\sc i} lines ranging from the near-IR to the near-UV.
CLOUDY photoionization equilibrium models 
including radiative transfer
are developed to predict the 
observed He~{\sc i} decrements
and the quantitative agreement is quite remarkable.
Following from these results, select He~{\sc i}
lines are combined with H~{\sc i} and [O~{\sc ii}] lines and stellar 
extinction data to validate a new normalizable analytic expression for the
wavelength dependence of the extinction.
In so doing, the He$^{+}$/H$^{+}$ abundance is also derived.

\end{abstract}


\section{Introduction \label{intro}}
The Orion Nebula is the brightest and best-resolved H~{\sc ii} region
and is the defining case with a blister geometry \citep{zuc73,bal74}.
Our study of this region
makes use of emission-line diagnostics from HST FOS and STIS
spectra supplemented by
extensive ground-based
observations
\citep[e.g.,][]{bal96,rub97,bal00},
with the goal of
probing the nebula's physical structure and determining
a robust set of its elemental abundances.

To complete a full spectral analysis in the UV and visible,
we have had to develop and test an extinction curve amended slightly
from those often applied to visible Orion Nebula observations
\citep[][hereafter CP70 and CCM89, respectively]{cos70,car89}.
See \S~\ref{extinction_corrections}.
In the course of this work we examined the decrement
within various series of He~{\sc i}
lines, planning to use these lines to constrain extinction, primarily in the 
ultraviolet.
As reported in \citet{mar96} and now fully
described in \S~\ref{hei_decrement},
our FOS and STIS observations have revealed
a
decrement within the ultraviolet $2~^3S-n~^3P$ series that is not in
accord with case~B \citep[as defined in][]{bak38} recombination theory 
\citep{smi96, ben99, por05},
and cannot be explained by extinction effects.
This anomalous decrement 
stems from the metastability of $2~^3S$,
leading to radiative transfer effects which then affect other series 
\citep{rob68,ost89}.
We compare the predictions of this
theory with our set of
HST FOS and STIS observations -- the first detailed
observational analysis of
He~{\sc i} UV lines originating from high $n$ terms.
In a comprehensive examination including other datasets we also show
quantitatively how the same theory self-consistently accounts for resonance 
fluorescence enhancement observed in several other lines of two related series.
Including the radiative transfer effects
reduces $\chi^2$ by a factor of roughly 10 as compared to case~B recombination alone,
showing a remarkable agreement between radiative transfer theory and observations.
%
\section{Observations}
\begin{figure}
\epsscale{0.8}
\centerline{
\rotatebox{270}{
\scalebox{1.0}[1.0]{
\plotone{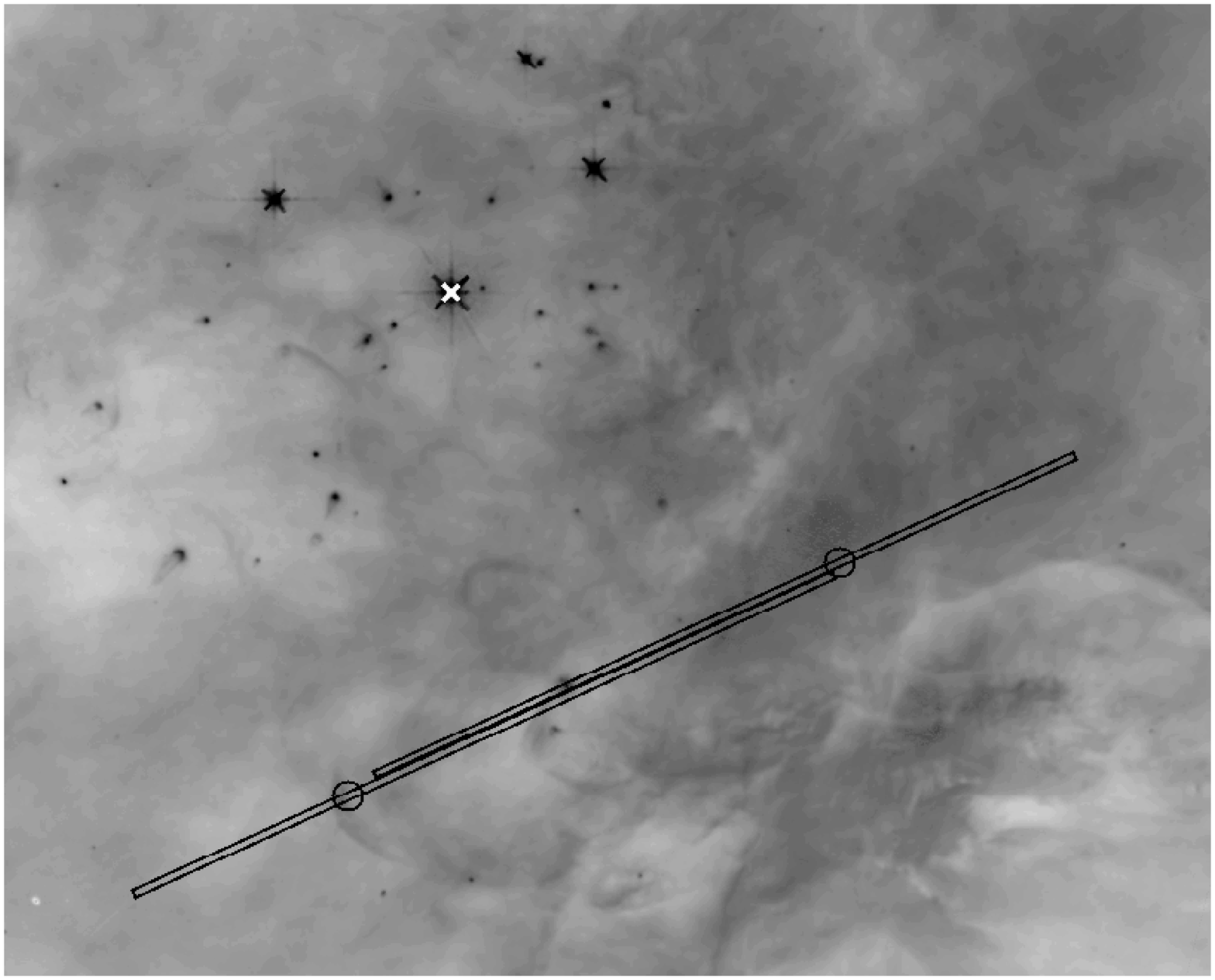}
}}
}
\caption[STIS Slits~1 and 2 and FOS 1SW and x2 positions]{HST STIS Slit~1 
($52\arcsec\times0.5\arcsec$, right) and Slit~2 (left) shown overlapping 
with HST FOS positions 1SW (right 
circle) and x2 (left circle) \citep{bal96, rub97}, respectively.  Slit positions
are overlaid on F656N (H$\alpha$) WFPC2 image.  $\theta^1$~Ori~C is marked with a 
$\times$. N is up and E to the left.  
\label{slits1and2}}
\end{figure}

HST FOS and STIS spectra were obtained with spectral coverage from the UV
(1600\AA) through to the visible (7400\AA) for two lines of sight
\citep[roughly 1SW and x2,][]{bal96,rub97}.
The FOS aperture has a diameter $0\farcs86$ and
the STIS slit is $52\arcsec\times0\farcs5$ (refer to Fig.~\ref{slits1and2})
of which only (the roughly central)
$28\arcsec$ are recorded using
the STIS UV detectors (FUV-MAMA, NUV-MAMA).
The UV and optical spectra were adjusted to align spatially using a bright
feature common to all spectra: proplyd 159-350 \citep{ode94}.  

The placement of Slit~1 (see Fig.~\ref{slits1and2}) was such that it was centered on a region
of the nebula where there is an extinction gradient. 
Because of this, we divided the $52\arcsec\times0\farcs5$ slit into four 
$13\arcsec\times0\farcs5$
sections\footnote{All STIS spectra have small fiducial bar sections which interfere with the spatial 
coverage 
of the spectra.  For this reason, when average 
surface brightnesses of tiles are being calculated along 
the slit, these fiducial bar sections are avoided.}, labeled ``a'' through ``d''.
The different spatial coverage of the MAMA and CCD detectors means that the outer two 
sections (SLIT1a and SLIT1d) have reliable coverage only in the visible, but
the central two sections (SLIT1b and SLIT1c) have coverage in the visible and the UV.
The same sectioning was done for Slit~2 (x2, see Fig.~\ref{slits1and2}).  However,
there was little change in the extinction across Slit~2 so the two central
sections (SLIT2b and SLIT2c) should be similar.

The reduction of STIS spectra was
performed as
in \citet{rub03} and the reduction of FOS spectra as in \citet{rub98}.
Our HST STIS (SLIT1b, SLIT1c, SLIT2b, SLIT2c) and FOS (1SW) observations are presented in 
Table~\ref{uncorrected_table}.

A wavelength-dependent extinction correction must still be applied to the entire spectrum
(UV through visible/near-IR).  To date, there has not been a robust way of reliably correcting
such a large portion of the spectrum to allow for full (i.e., UV through visible) spectral analysis.
However, with the broad spectral coverage of the HST data, we are able to
develop and test a smooth consistent extinction correction curve.
The development and validation of such an analytic extinction curve is
discussed in \S~\ref{extinction_corrections}.
To ensure that any extinction curve is 
well constrained,
we must maximize the number of lines
used in its determination.  In \S~\ref{hei_decrement}, we discuss the
basis for the inclusion of He~{\sc i} (especially 
He~{\sc i} UV) lines.

\section{Extinction corrections \label{extinction_corrections}}
Were it not for extinction, the Orion Nebula would appear much brighter.
Typical optical depths reported are $\tau_V\sim1$ .
From the shape of the extinction curve
\citep[e.g., see curves in][]{kim94},
the optical depth in the Lyman continuum ($\tau_{912}$) would be
several times this value.
The relatively soft radiation from the primary ionization source (O star, $\theta^1$~Ori~C) 
cannot photoionize an optical depth to dust of much more than 
$\tau_{912}\sim1$ indicating that
a large part of the optical extinction must come
from neutral material in the foreground.
Further evidence for and properties of this ``veil'' are summarized by 
\citet{abe04}.

All observed emission-line surface brightnesses or fluxes, from the
infrared to ultraviolet, are affected by this wavelength-dependent
extinction and must be corrected before further analysis and
interpretation.  Extinction is characterized by its wavelength
dependence (shape) and the amount at a given
wavelength (amplitude).

Stellar extinction deduced from 
stars in the
nebula provides a useful reconnaissance of the shape of the 
Orion Nebula extinction
curve, providing a continuous curve with which to make
predictions (by interpolation) for any given line observed.
We have reconsidered the
wavelength dependence of the infrared,
optical and ultraviolet extinction, presenting our results as 
a convenient analytic
expression describing this shape (``normalized extinction curve'')
in \S~\ref{shape}.
The amplitude by which to scale the normalized extinction curve can
be determined for a given line of sight from nebular lines whose ratios of
line emissivities
are known from atomic theory (e.g., H~{\sc i} Balmer series).  An optimal 
implementation is discussed in \S~\ref{amplitude}.
In \S~\ref{comp_cp}, we compare our extinction curve with the nebular 
extinction curve of CP70.
In \S~\ref{red_validation}, we discuss our extensive validation of
the use of our stellar-based extinction curve for
interpolating corrections to the nebular emission in the 
optical, the ultraviolet
 and the near-infrared.

\subsection{Parameterization of extinction \label{parameterization}}
We have already stressed how one requires a shape ($s_{\lambda}$) and
amplitude ($a$) to describe the extinction:
\begin{equation}
\tau_{\lambda} = a\, s_{\lambda}.
\end{equation}
For different lines of sight, grain properties like size distribution
and composition are manifested in $s_{\lambda}$, while $a$ scales with
the column density of dust.

Let $F$ represent the observed flux of a patch of nebula (surface
brightness times solid angle) or of a star, with subscript $o$ indicating
the intrinsic flux -- as it would have been had it not been affected by
extinction $\tau$. (The dependence on wavelength $\lambda$ has been
suppressed.)  Thus $F/F_o = \exp(-\tau)$.  For a uniformly bright
nebular patch this is also the ratio of the specific intensities,
$I/I_o$.  Note that $I_o$ can be predicted by atomic theory, from the
emissivity $j$ and (for most lines) the emission measure
\citep{ost89}. We adopt $I_o \propto j$ in the applications here, so
that
\begin{equation}
F_{\lambda} = {\rm constant} \times j_{\lambda} \; \exp(-a\, s_{\lambda}).
\label{fit}
\end{equation}

The logarithmic (base 10) extinction is $C = -\log(F/F_o) =
-\log(\exp(-\tau)) = 0.4343 \tau$.  The extinction in magnitudes is $A =
-2.5\log(F/F_o) = 2.5 C = 1.086 \tau$. Given the linear transformations
between $A$, $C$, and $\tau$, a ratio like $A_{\lambda_1}/A_{\lambda_2}$
evaluated for two wavelengths is clearly interchangeable with the same
ratio for $C$ or $\tau$.

In the literature there are various parameterizations of $A$, $C$, and
$\tau$ that are worth distinguishing, since they imply different choices
and normalizations of $s_{\lambda}$. 
\citet{ost89} uses $s_{\lambda} = 0.5
\tau_{\lambda}/(\tau_{\mathrm{H}\gamma} -\tau_{\mathrm{H}\alpha})$ and comments on the  
utility of the differential form $s_{\lambda} - s_{{\mathrm H}\beta}$.
Note that he uses the notation $f_{\lambda}$ where we use
 $s_{\lambda}$.

For nebular extinction, one often writes the logarithmic extinction as
$C_{\lambda} = C_{{\mathrm H}\beta} (1 + f_{\lambda})$, with $f_{{\mathrm 
H}\beta} = 0$
\citep[e.g.,][]{pei77}. 
Then if we adopt $s_{\lambda} = (1 + f_{\lambda}) \equiv
\tau_{\lambda}/\tau_{{\mathrm H}\beta}$ we have $a = C_{{\mathrm H}\beta}/0.4343 =
\tau_{{\mathrm H}\beta}$.

For stellar extinction one often
adopts the normalization $s_{\lambda}=A_{\lambda}/A_V \equiv
\tau_{\lambda}/\tau_V$
(e.g., CCM89)
in which case
$a = A_V/1.086 = \tau_V$.  In order to compare these differently 
normalized nebular and stellar extinctions, one can easily show that
\begin{equation}
1 + f_{\lambda} = \frac{A_{\lambda}/A_V}{A_{{\mathrm H}\beta}/A_V}.
\end{equation}

Differential extinction between two wavelengths can of course be defined
in any system.  For example, color excesses are differences in
magnitudes, like $E_{B-V} = A_B-A_V$. These can lead to differential
extinction curves with new normalizations, like $E_{\lambda-V}/E_{B-V} =
(A_{\lambda}-A_V)/E_{B-V} = R_V(A_{\lambda}/A_V-1)$, accompanied by the 
introduction of new parameters, like the ratio of total to selective    
extinction, $R_V = A_V/E_{B-V}$.

\subsubsection{Determining the shape, $s_{\lambda}$: Stellar extinction
\label{shape}}
CCM89 defined the shape of optical stellar extinction
analytically using a single parameter, $R_V$, to describe differences
between lines of sight.
The shape has
been adopted in many investigations of the Orion Nebula
(Baldwin et al.\ 1991; Osterbrock et al.\ 1992, hereafter OTV92; Greve et al.\ 1994),
with $R_V = 5.5$.\footnote{$R_V = 3.1$ in the diffuse
interstellar medium \citep{sav79}.}

\begin{figure}
\epsscale{1.0}
\centerline{
\rotatebox{270}{
\scalebox{1.0}[1.0]{\plottwo{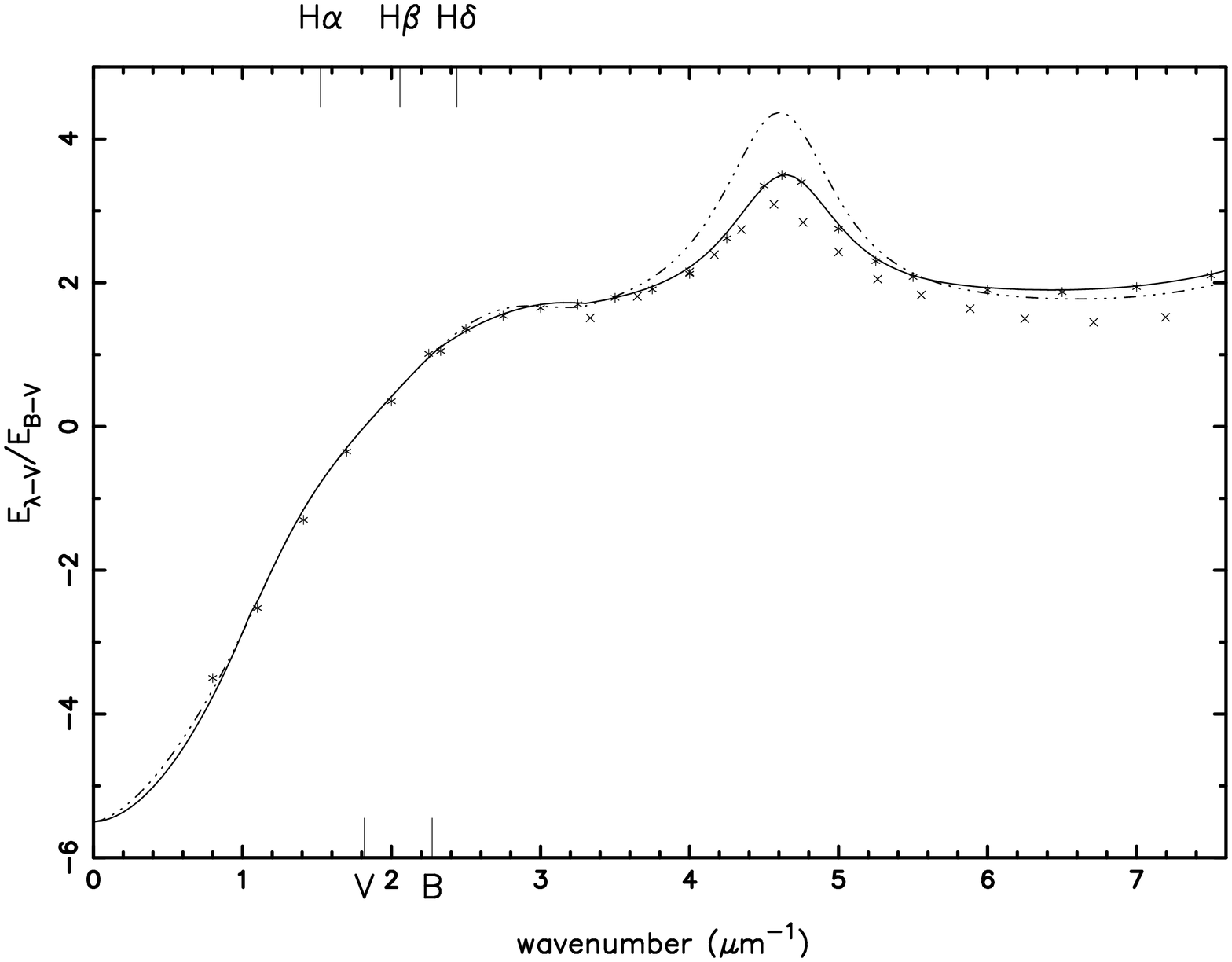}{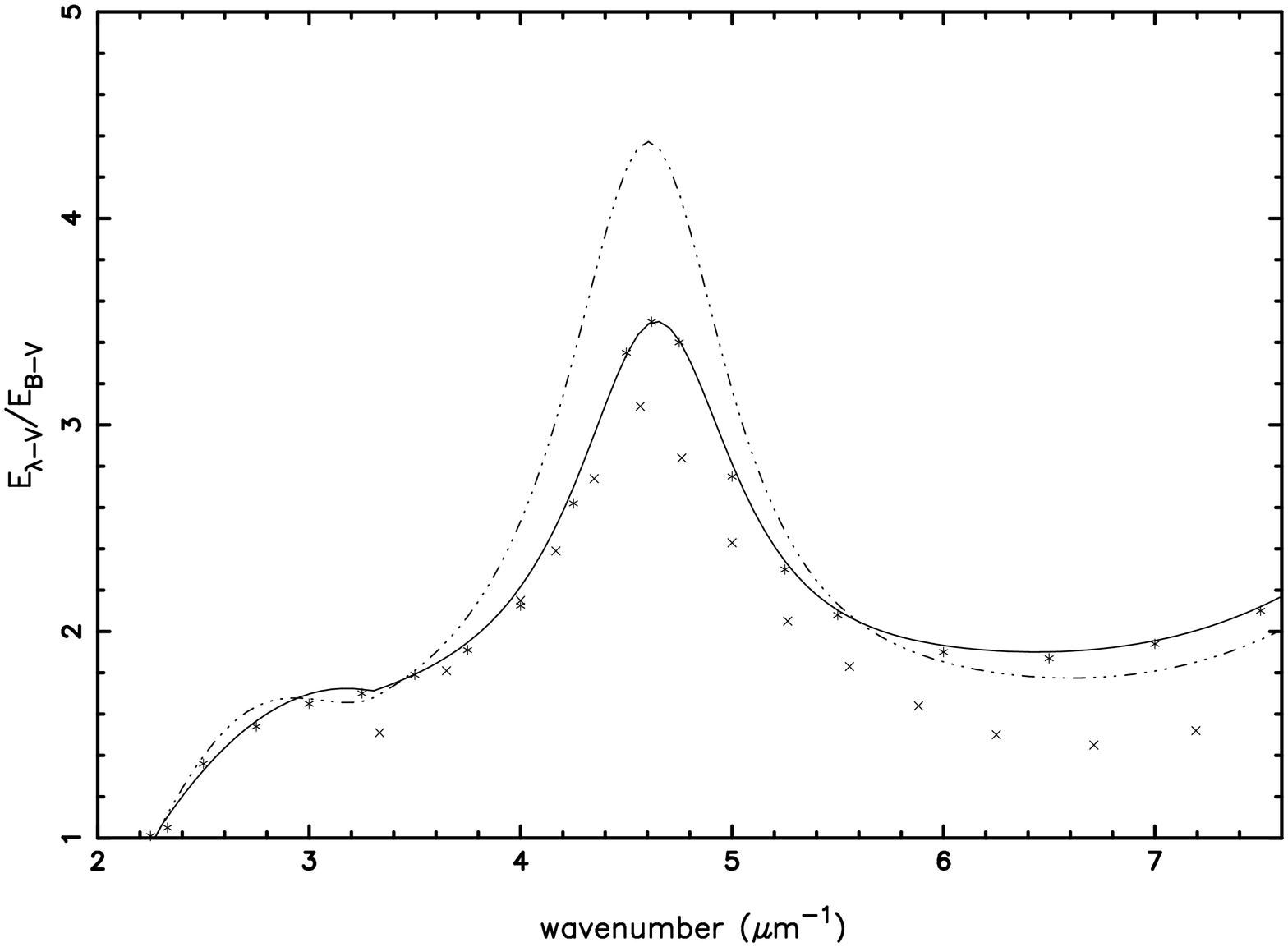}}}}
\caption[UV and IR modifications made to CCM89 formulation]{
Analytic stellar-based extinction curves.
$(top)$ Dash-dotted line:
CCM89  extinction curve  presented here in differential form, $E_{\lambda-V}/E_{B-V}$ ($R_V=5.5$) for ease of
comparison with original BS81 data ($\times$).
BS81 data should be adjusted by roughly +0.25 mag -- the result of a calibration
error
pointed out by CC88, whose corrected stellar data are also plotted ($\ast$).
Solid line: our modification of the CCM89 formulation to better fit the CC88 corrected data
and shape near 2.7~$\mu$m$^{-1}$. $(bottom)$ Same as above, but the window has been scaled to
accentuate the differences between CCM89 and our modified curve.
\label{fig4}}
\end{figure}

\subsubsubsection{Optical}
Good optical spectrophotometry of the Orion stars ($\theta^1$ and
$\theta^2$) is presented by
\cite{car88} (hereafter CC88)
as $E_{\lambda-V}/E_{B-V}$.
Using measurements from their graphs, we show that the optical
data (and more scanty infrared data) are fit well by the CCM89 formula
(dash-dotted line in our Fig.~\ref{fig4}).
The CC88 data (asterisks in Fig.~\ref{fig4})
continue to about 3~\micron$^{-1}$ and show that there is a
knee in the CCM89 curve at about 2.7~\micron$^{-1}$ ($\lambda$3700) that is
perhaps a bit too high and sharp.
Therefore, an analytic modification\footnote{
CCM89's equations~3a and 3b are modified to
$a(x) = 1.192528 + 0.27592(x - 2.3) - 0.15733(x - 2.3)^2$ and $b(x) = 0$, but only for
2.3~\micron$^{-1} < x < 3.3$~\micron$^{-1}$.
} has been made to the CCM89 formula
(solid line) to round off the knee smoothly between 2.3 and
3.3~\micron$^{-1}$, fitting the CC88 data  quite well.

\subsubsubsection{Ultraviolet}
The ultraviolet data come from IUE measurements by \citet{boh81}
(hereafter BS81) for the same Orion stars (see Fig.~\ref{fig4}).  
The BS81 points (x) do not join on
smoothly to the CC88 optical data.  CC88 reanalyzed the ultraviolet data,
showing how it could reasonably be joined
smoothly to their new accurate optical spectrophotometry (asterisks).
The main effect on the BS81 data (x)
is to move it up slightly ($\sim0.25$~mag shift), concomitantly increasing 
the
expected ultraviolet extinction (see Fig.~\ref{fig4}).

Note that in the ultraviolet, the analytic CCM89 formula  does
not fit either the CC88 or BS81 observations: the predicted bump is too
strong. However, it is possible to adjust a few constants\footnote{
CCM89's equation~4b becomes: $b(x) = -2.9 + 1.825x + 0.93/[(x - 4.65)^2 + 0.263]+F_b(x)$
for 3.3~\micron$^{-1} < x < 8$~\micron$^{-1}$.
} in the CCM89
formula to obtain a modified CCM89 curve (solid line) that runs smoothly
through the CC88 data.

To gauge the importance of these modifications for the determination of
extinction-corrected fluxes, consider the situation for the N~{\sc ii}]
$\lambda\lambda$2140-43 pair near the peak of the extinction curve.  The
unmodified
CCM89 curve exceeds ours by $\delta  E_{\lambda-V}/E_{B-V} = 0.85$.
For a typical value $E_{B-V} = 0.3$ the difference would be $\delta \tau =
0.23$; thus the N~{\sc ii}] line would be 26\% (0.1~dex) stronger if corrected
according to the unmodified CCM89 curve.

\subsubsubsection{Near infrared \label{ext_nearIR}}
Near 1~\micron\ the CCM89 $A_{\lambda}/A_V$ curve becomes a power law in
$\lambda^{-1}$ of index 1.6.  Most stellar curves exhibit this ``common
power law'' region of the extinction curve, independent of $R_V$
\citep{mar90}. The latter analysis suggests an index of 1.85, which we
adopt in our extinction curve.  This modification will not
have an impact here,
as the observations analyzed do not extend much beyond 1~\micron.

We have included a table of $f_{\lambda}$ values for a number of observed emission
line wavelengths (see Table~\ref{flambdatab}),
calculated directly from our modified set
of CCM89 equations (with $R_V=5.5$).  For other wavelengths,
refer to CCM89 and our analytic modifications discussed above.

\subsubsection{Setting the amplitude, $a$: Optimal fitting of emission 
lines
\label{amplitude}}
Measurements of emission lines in nebulae are often used to gauge the
extinction
(e.g., CP70; OTV92; Greve et al.\ 1994; Esteban et al.\ 1998, hereafter EPTE98; Esteban et al.\ 2004, hereafter EPG04).
If
the shape $s_{\lambda}$ of the extinction curve is known, then the
amplitude $a$ in equation \ref{fit} can be determined with even a single
line pair from the same ion, assuming the relevant emissivities
$j_{\lambda}$ are known.  The Balmer and Paschen lines are popular as
their emissivities are well described by H~{\sc i} recombination theory and
can be computed readily for various electron temperatures ($T_e$) and
densities ($N_e$).  See \citet{sto95} for detailed access to the results
of their computations using an interactive data server.  For the
illustrations below we adopt case~B emissivities with $T_e = 8000$~K, $N_e = 2500$~cm$^{-3}$
for FOS-1SW and STIS-SLIT1c (as determined from nebular models presented in 
\S~\ref{cloudy_extmodels}),
$T_e = 9000$~K, $N_e = 5000$~cm$^{-3}$ for the remainder of the STIS slits,
and 
$T_e = 8300$~K, $N_e = 8900$~cm$^{-3}$ for the EPG04 observations
(as determined from their temperature and density diagnostic ratios).
Lines from other ions can be useful too  
(e.g., He~{\sc i}, [O~{\sc ii}], [S~{\sc ii}]), as further illustrated in 
\S~\ref{hei_decrement} (He~{\sc i}) and 
\S~\ref{red_validation} ([O~{\sc ii}], [S~{\sc ii}], He~{\sc i}).

\subsubsubsection{Method}
Sometimes the amplitude is taken from only one line ratio, like
\halpha/\hbeta, but often other lines are available too.  Past 
methodology has been to form various line ratios, compute the amplitude
(e.g., the equivalent $A_V$) from each independently (sometimes with   
discrepant results), and then compute some average.  A new approach    
adopted here is to fit the observed line fluxes directly to equation   
\ref{fit}, not forming line ratios at all.  The fit is carried out     
straightforwardly by non-linear least squares, with two unknowns: $a$  
and a constant multiplier (which is proportional to the ionic abundance
through the emission measure.)
In addition, from the goodness of fit of the data to the model, we
obtain the $1\sigma$ confidence intervals on the parameters.
This    
approach has some advantages: avoiding the inevitable bias from errors in the 
chosen normalizing line(s), using a standard treatment of the measurement
errors to weight the fit and hence avoiding the ambiguity associated with
deciding what average amplitude to compute.

If later one wants to tabulate extinction-corrected line ratios relative
to, say, H$\beta$ (or He~{\sc i}~4471 if He~{\sc i} lines are used in the extinction fitting),
then the extinction model gives the 
best estimate of   
the corrected  flux for this normalization, optimally consistent 
with all other lines in the series.  When the He~{\sc i} recombination lines are fit simultaneously,
then the optimal He$^{+}$/H$^{+}$ ratio is obtained, along with its formal error, without reference
to particular He~{\sc i} to H~{\sc i} dereddened line ratios.

\subsection{Comparison between stellar and nebular extinction curves \label{comp_cp}}
The CP70 nebular extinction curve
 is an alternative commonly used in the
calculation of the extinction correction (EPTE98; EPG04).
This curve was determined from optical and radio observations of four
lines-of-sight in the Orion Nebula.  The radio (1.95~cm) observations are used to
determine
the absolute extinction, $A_{{\mathrm H}\beta}$, and the piecewise normalized 
extinction curve, $1+f_{\lambda}$, is obtained for various lines:
\begin{equation}
f_{\lambda} = \frac{E_{\lambda-{\mathrm H}\beta}}{A_{{\mathrm H}\beta}}.
\label{cpflambda}
\end{equation}

Calculations of $C_{{\mathrm H}\beta}$ offer a 
simple means by which to compare our modified CCM89 curve with that of CP70.
Performing
a least-squares analysis on EPG04 data using the CP70 curve (and 
interpolations thereof),
we get $C_{{\mathrm H}\beta} = 0.65\pm0.03$, as
compared to $C_{{\mathrm H}\beta} = 0.85\pm0.04$ when we use our extinction curve.
EPG04 had $C_{{\mathrm H}\beta} = 0.76\pm0.08$ when using this same series 
(up to H15, 
$\lambda$3712) of 
H~{\sc i} lines.  However, this comparison
of $C_{{\mathrm H}\beta}$ is not enough.
The shape of the CP70 curve is investigated more thoroughly in the following sections.

\subsubsection{Infrared discrepancy \label{IRdiscrepancy}}
\begin{figure}
\epsscale{0.7}   
\rotatebox{270}{
\centerline{\scalebox{1.0}[1.0]{\plotone{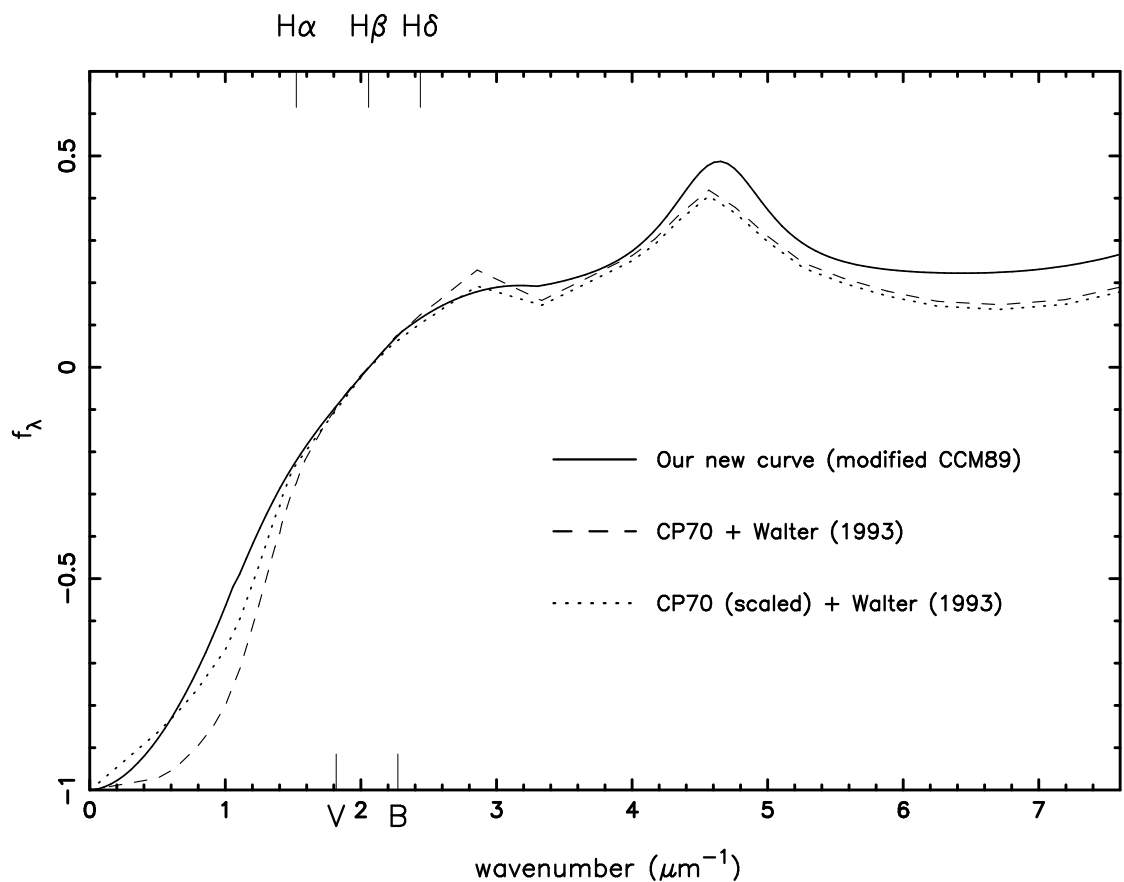}}}}
\caption[Extension of the CP70 curve into the UV and the renormalized CP70 curve]{ CP70 extinction curve (dashed line)
including extension into the UV using uncorrected BS81 data following \citet{wal93}.
Our new extinction curve (solid curve) is
overplotted, showing an unexplained discrepancy in the IR portion of the CP70 curve.
The renormalized
CP70 nebular extinction
curve (dotted line) shows a marked improvement to the IR discrepancy.  The UV portion of the scaled CP70 curve is rederived directly
from the BS81 $E_{\lambda-V}/E_{B-V}$ data, using the new scaled $f_V$ and $f_B$.
\label{waltcp}}
\end{figure}

The zero-point is calculated by CP70 using radio data at
(more or less) the same spatial position as the optical data.   
The rest of CP70's $f_{\lambda}$ is
determined directly
from specific Balmer or Paschen lines \citep[as tabulated in][]{pei77},
so that
most of the curve plotted in Figure~\ref{waltcp} is interpolated
\citep[by CP70;][]{pei77,wal93}.

Near 1~\micron$^{-1}$, the CP70 curve is dramatically below our extinction curve
(see Fig.~\ref{waltcp}).
This is getting into the
``common power law'' region of the extinction curve mentioned in
\S~\ref{ext_nearIR}.
Nowhere is there evidence for the strong curvature implied
in the CP70 extinction curve in the region from 1~\micron$^{-1}$ to
zero wavenumber.
The IR discrepancy might simply be the result of 
an inappropriate interpolation of the CP70 extinction curve.
Looking at Figure~1 of
CP70, it appears as if other interpolations to the $V$ band are
possible.  
If the interpolated $f_V$ were a bit more negative ($-0.12$ or $-0.13$ instead of
$-0.094$ would seem reasonable) and
the CP70 curve (shown in our Fig.~\ref{waltcp}) was refit (and rescaled, see below),
it would be noticeably closer to the stellar one from the optical through
the near infrared.

The change in the interpolated $f_V$ would affect $R_V$
as derived from
\begin{equation}
R_V = \frac{1+f_V}{f_B-f_V},
\end{equation}
resulting in $R_V = 4.5$ ($f_V=-0.12$) or 4.2 ($f_V=-0.13$) compared to the CCM89 and CP70
value of $R_V = 5.5$.
However, 
the $f_{\lambda}$ 
depend on
 comparing radio and optical data (see equation~\ref{cpflambda}).
This radio/optical comparison is almost certainly
affected by a 10-20\% reflection effect in the optical \citep[e.g.,][]{hen98}.  Because of reflection,
the lines like \hbeta\ are measured to be too bright relative to the
radio,
resulting in a low estimate of the 
absolute extinction, $A_{{\mathrm H}\beta}$.
In the optical where the
differential extinction $E_{\lambda-{\mathrm H}\beta}$ is
not affected much by reflection, 
$f_{\lambda}=E_{\lambda-{\mathrm H}\beta}/A_{{\mathrm H}\beta}$
will be too high.
A typical logarithmic extinction $C_{{\mathrm H}\beta}$
($C_{{\mathrm H}\beta}=0.4 A_{{\mathrm H}\beta}$) is 0.5, so that 
with $f_V = -0.12$, the logarithmic extinction at $V$ is $0.5 \times (1 - 0.12) =
0.44$ or $A_V\sim1$.  A 20\% reflection effect at $V$ would cancel
out real extinction at the level $\Delta~A_V= 0.2$, lowering what should be
1 to 0.8 (apparent). Then $0.8 \times 5.5$ predicts $R_V$(apparent)$
= 4.4$.  Both the sign and order of magnitude of the reflection
correction appear to be right.
Note that there are other possibilities of systematic error as
the radio/optical comparison also depends on the 
adopted temperature, matching the spots observed, etc.

To remedy this artificial alteration of $R_V$,
we introduce a correction to $A_{{\mathrm H}\beta}$ -- effectively
a rescaling of $f_{\lambda}$ -- to maintain $R_V=5.5$.
With $f_V = -0.12$ in our new interpolation of the CP data, one needs to divide
all $f_{\lambda}$ by a factor 1.198 to ensure that $R_V = 5.5$.
This rescaled CP70 absolute extinction curve
is presented in Figure~\ref{waltcp}.

The shape of our rescaled CP70 curve in the visible is similar to that
of our modified CCM89 curve and so regardless of which of these is used, one should be able to get good 
differential 
fits to a series of optical lines.  The differential
extinction between \halpha\ (1.52~\micron$^{-1}$) and \hdelta\
(2.44~\micron$^{-1}$)
is the same in both curves, and one  
obtains more or less the same $C_{{\mathrm H}\beta}$
when
fit to the Balmer lines,
independent of the choice of
curve (see \S~\ref{validoptical}).

\subsubsection{Ultraviolet \label{stell_nebUV}}
\citet{wal93} extended the CP70 curve to the ultraviolet  by
attaching on the BS81 data assuming $R_V = 5.5$.
This extended extinction curve was
also described by
\citet{rub93}.
We have derived $f_{\lambda}$ values from the original BS81 
$E_{\lambda -V}/E_{B-V}$ form
\citep[as in][]{rub93}
for both the CP70 and rescaled CP70 curves, generating
a shape of the ultraviolet extinction for each, as seen in Figure~\ref{waltcp}.
This extension lies below the
CC88 revision of the BS81 data in the ultraviolet and does not join smoothly to the optical
portion of the CP70 curve.  One of the reasons is that the CP70 curve has
been extrapolated to $\lambda$3500 (see their Table 2), incorrectly high,
beyond the highest Balmer transitions that were observed (H9,
$\lambda$3835).  When combined with the ultraviolet data (BS81 version)
this really sticks out as a kink in the curve at
$\lambda$3500; a smooth join, like all (stellar) extinction curves
observed, is not possible. The reddening correction of a line like
[O~{\sc ii}] $\lambda$3727 would be in doubt.  Our modified CCM89 curve seems a
more reliable alternative in this ``extrapolated'' region.

\subsection{The stellar extinction curve: validation \label{red_validation}}
It is important to address whether the shape of extinction
derived from stars is appropriate for diffuse nebular emission, where
the presence of scattered light might be an issue.
We have therefore
carried out extensive tests of this technique using consistent sets of
nebular measurements of hydrogen lines in Orion, both from the
literature
(CP70; Peimbert \& Torres-Peimbert 1977; OTV92; EPTE98; EPG04)
and from 
our HST and ground-based 
\citep[Baldwin et al.\ 2000, hereafter BVV00;][]{bla06a,bla06b} 
observations.

Here we exploit our HST FOS and STIS (SLIT1c) observations for the line of sight 1SW
(see Table~\ref{uncorrected_table})
as,
with the addition of several UV lines,
they offer an extensive spectral range
for the 
validation.
As will be discussed here, and again in \S~\ref{ext_corr_revisited} (Figures~\ref{FOSfit} and 
\ref{STISfit}), the data are well fit using our extinction curve.

\subsubsection{Optical \label{validoptical}}
\begin{figure}
\epsscale{1.0}
\centerline{
\rotatebox{270}{
\scalebox{1.0}[1.0]{\plottwo{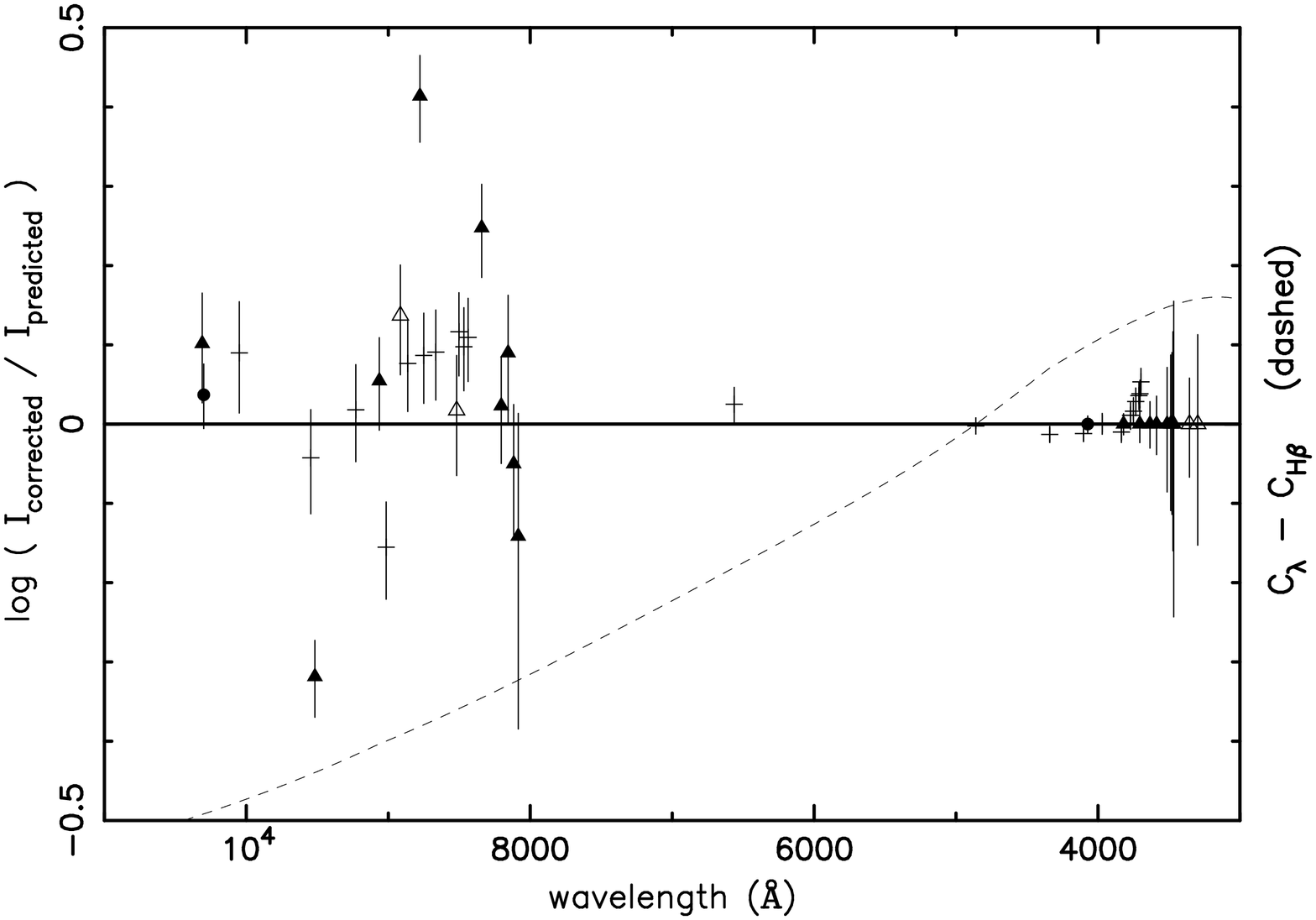}{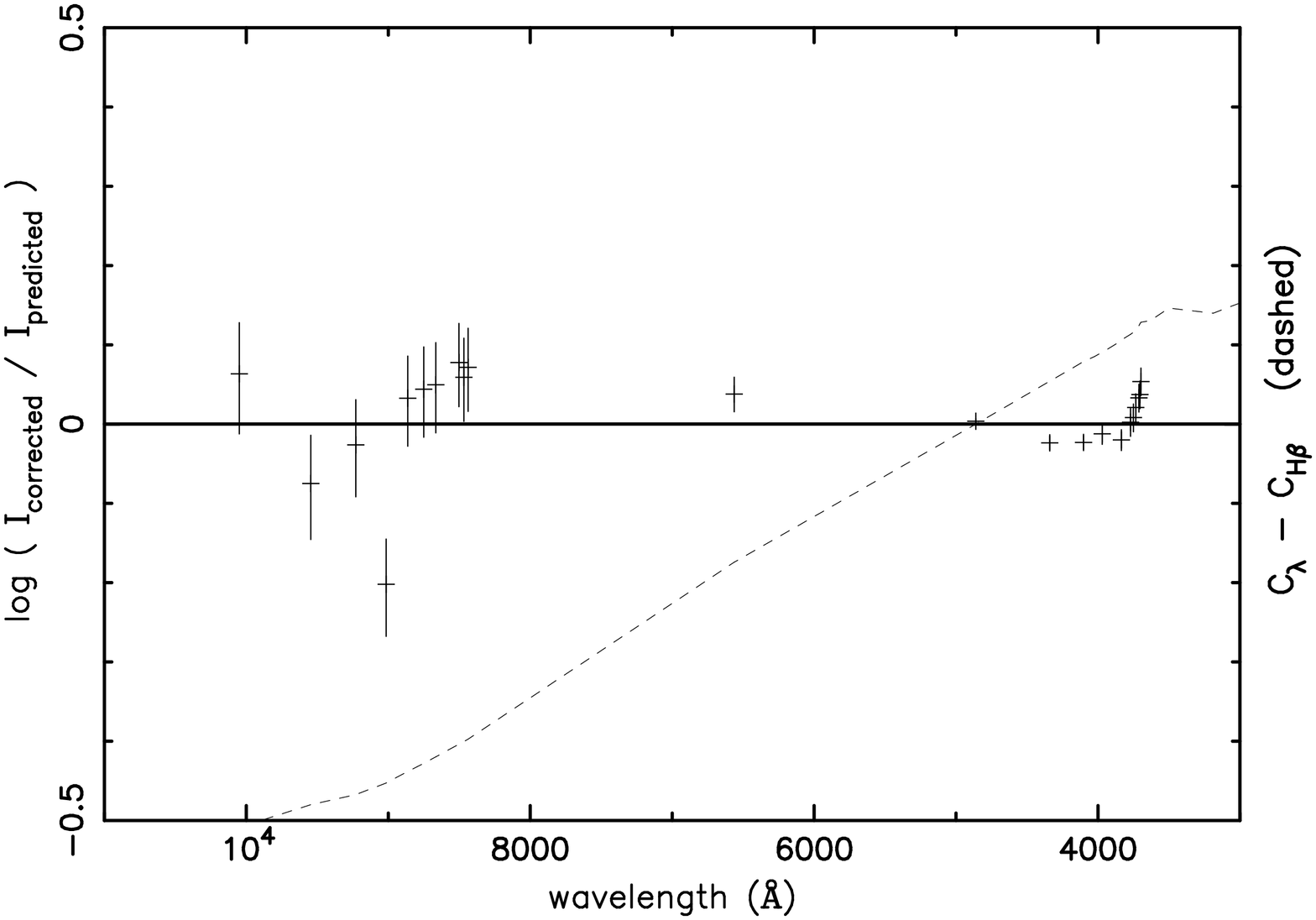}}}}
\caption[Our analytic extinction curve
and
CP70 curve as determined from
EPG04 observed data]{ Best-fit differential extinction curves
(dashed line) as determined from
EPG04 H~{\sc i} Balmer and Paschen lines.
Residuals and
error bars are plotted
for each H~{\sc i} emission line (+)
used in the fit.
$(top)$ our modified CCM89 extinction curve, $R_V=5.5$,
$C_{\mathrm{H}\beta}=0.82\pm0.04$.  Also shown are data from common
upper-level pairs
 of [S~{\sc ii}] ($\bullet$) and He~{\sc i} triplets ($\blacktriangle$) and singlets ($\triangle$)
after applying this reddening correction. The optical member of the pair, usually with the lower error,
has been plotted as $I_{corrected}=I_{predicted}$.
$(bottom)$ Residuals for extinction correction according to our scaled CP70 nebular curve, $C_{\mathrm{H}\beta}=0.76\pm0.04$.
The common-upper level pairs are not included, but would result in a similar scatter about log($I_{corrected}/I_{predicted}$)$=0$.
\label{est_ccm}}
\end{figure}

In addition to our own STIS and FOS observations we have used others' observations to examine
the fit of the stellar-derived extinction curve in the optical.
Using the extensive observations of EPG04,
we have determined the best-fit extinction curve
(see 
\S~\ref{shape}) 
using the unblended H~{\sc i} Balmer lines (up to H17, $\lambda$3697) and Paschen lines
(up to P18, $\lambda$8438),
resulting in
$C_{{\mathrm H}\beta} = 0.82\pm0.04$.
The resulting curve and residuals
from this fit
are shown in Figure~\ref{est_ccm}.
Over the spectral range that can be accessed by these
lines, the fit is good, demonstrating that 
our extinction curve provides a good empirically-derived interpolation formula for
differential extinction.

The EPG04 data were also fit using the scaled form of CP70 developed in 
\S~\ref{comp_cp}
and the same H~{\sc i} lines as above.
The resulting curve
and residuals are 
included in the bottom panel of Figure~\ref{est_ccm}.
We find $C_{{\mathrm H}\beta}=0.76\pm0.04$, in agreement with
$C_{{\mathrm H}\beta} = 0.76\pm0.08$ found by
 EPG04.
As shown in the top and bottom panels of Figure~\ref{est_ccm} and discussed in \S~\ref{comp_cp}, the
differential 
extinction is well represented by both our scaled CP70 and modified CCM89 curves.

We have also carried out fits of the various data sets using our
curve and $R_V = 3.1$, appropriate to the diffuse interstellar medium
\citep{sav79}.
By comparison to the Orion curve (with $R_V=5.5$) this curve is steeper throughout the
optical, producing relatively less (more) extinction compared to   
\hbeta\ in the near-infrared (blue).  This produces a markedly inferior
fit.

\subsubsection{Ultraviolet \label{validUV}}
Validation using nebular observations is especially important for the
ultraviolet portion of the extinction model.
The STIS and FOS spectra include the common upper level pair [O~{\sc ii}] $\lambda\lambda2471, 7325$
which can be used in this validation.

The [O~{\sc ii}] $\lambda$2471 line is actually a blend of
transitions from $^2P_{1/2}$ and $^2P_{3/2}$ to $^4S_{3/2}$ \citep{ost89}.
Transitions from these upper levels also lead to the formation of a pair of blended
near-infrared lines.
Each upper level contributes a line to a blend at $\lambda$7321
(transitions to $^2D_{5/2}$) and another line to a blend at $\lambda$7332
(transitions to
$^2D_{3/2}$); the combined near-infrared line will be denoted $\lambda$7325.
Thus the $I_{2471}/I_{7325}$ line ratio is not quite the ideal
case with a single common upper level in which the emissivities of the
line pair are simply proportional to the radiative transition
probabilities divided by the respective wavelengths.  Instead, this line
ratio depends on the collisional excitation of $^2P_{1/2}$ relative to
$^2P_{3/2}$. However, calculations with a five-level atomic model
show that the line ratio is not very sensitive to $T_e$ and $N_e$ over
the range expected in the Orion Nebula.
%
We adopt 0.75  using modern
collision strengths \citep{pra06} and older transition probabilities 
\citep{zei82}. We compare these results with a second ratio,
0.81 derived using the modern set of transition probabilities recommended by \citet{wie96}.
These are both calculated for $T_e=8000$~K and $N_e=2500$~cm$^{-3}$ (see \S~\ref{cloudy_extmodels}).

In Table~\ref{2471to7325},
the predicted $I_{2471}/I_{7325}$ ratios are compared to our observed ratio corrected using 
our extinction curve and H~{\sc i} lines.
Use of the modern transition probabilities results in a 15-20\% over-prediction, whereas the older
transition probabilities result in a slightly lower 10-15\% over-prediction.
These modern transition probabilities have been found to yield
anomalous
density and temperatures from [O~{\sc ii}] lines
\citep[see][EPG04]{wan04,bla06a}
and thus it is not unexpected that the older transition probabilities yield a
slightly more consistent result. 
%

As a check on the theoretical predictions, we have also examined the
[O~{\sc ii}] $I_{7321}/I_{7332}$ line ratio which is slightly more
sensitive to the relative upper level populations,
but not sensitive to the applied reddening correction. 
The prediction is
1.25 using \citet{zei82} transition probabilities and 1.23 using \citet{wie96} transition 
probabilities.
Our ratios
are 1.21-1.24 for the STIS observations.
Other observations in Orion by OTV92 give 1.134, while
BVV00 find 1.25 and EPG04 find 1.30.
Our ability to accurately predict (within 15\%) both the $I_{7321}/I_{7332}$ line ratio and
the common-upper-level line 
ratio
$I_{2471}/I_{7325}$ supports
the validity of our analytic extinction curve in the UV.
However, it would be beneficial to have further constraints in the UV.  This is addressed in
\S~\ref{hei_decrement}.


\subsubsection{Near infrared \label{valid_IR}}
Similarly, the [S~{\sc ii}] common upper-level line pair $I_{10300}/I_{4072}$ can be used to confirm the
validity of the extinction curve in the infrared.  Our data do not extend into the infrared, but
those of EPG04 do.  The three observed lines at $\lambda10300$ and two at 
$\lambda4072$ have 
been corrected for extinction using our
best-fit 
extinction curve
(see \S~\ref{validoptical}).
Since this is a ratio being used for validation and is not in the fit,
the visible line (which tends to have the lower error) has been set to $I_{corrected}/I_{predicted}=1$ in Figure~\ref{est_ccm}.
The agreement is good.

As pointed out in \citet{por06}, EPG04 have also observed a series of 
He~{\sc i} common upper-level line pairs.
These line pairs
have been overplotted on Figure~\ref{est_ccm} too.
The IR pair members have $I_{corrected}/I_{predicted}=1.2\pm0.5$.
While there are large differences between the observed and predicted line strengths,
there is no significant systematic bias, 
indicating that our
extinction curve is also valid between the near-IR and near-UV.

\section{Anomalous He~{\sc i} decrements \label{hei_decrement}}
\begin{figure}
\epsscale{1.0}
\centerline{
\rotatebox{270}{
\scalebox{1.0}[1.0]{
\plottwo{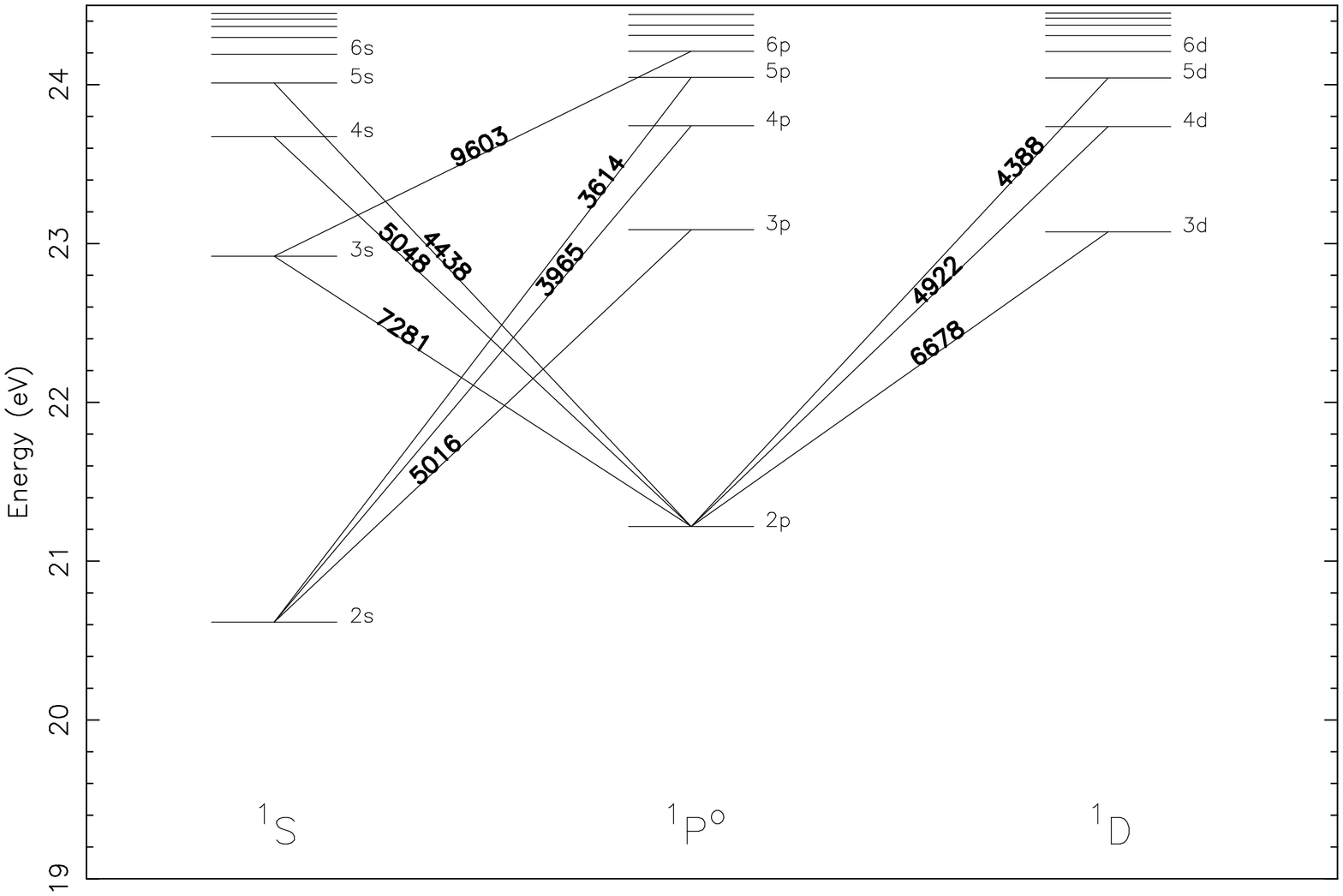}{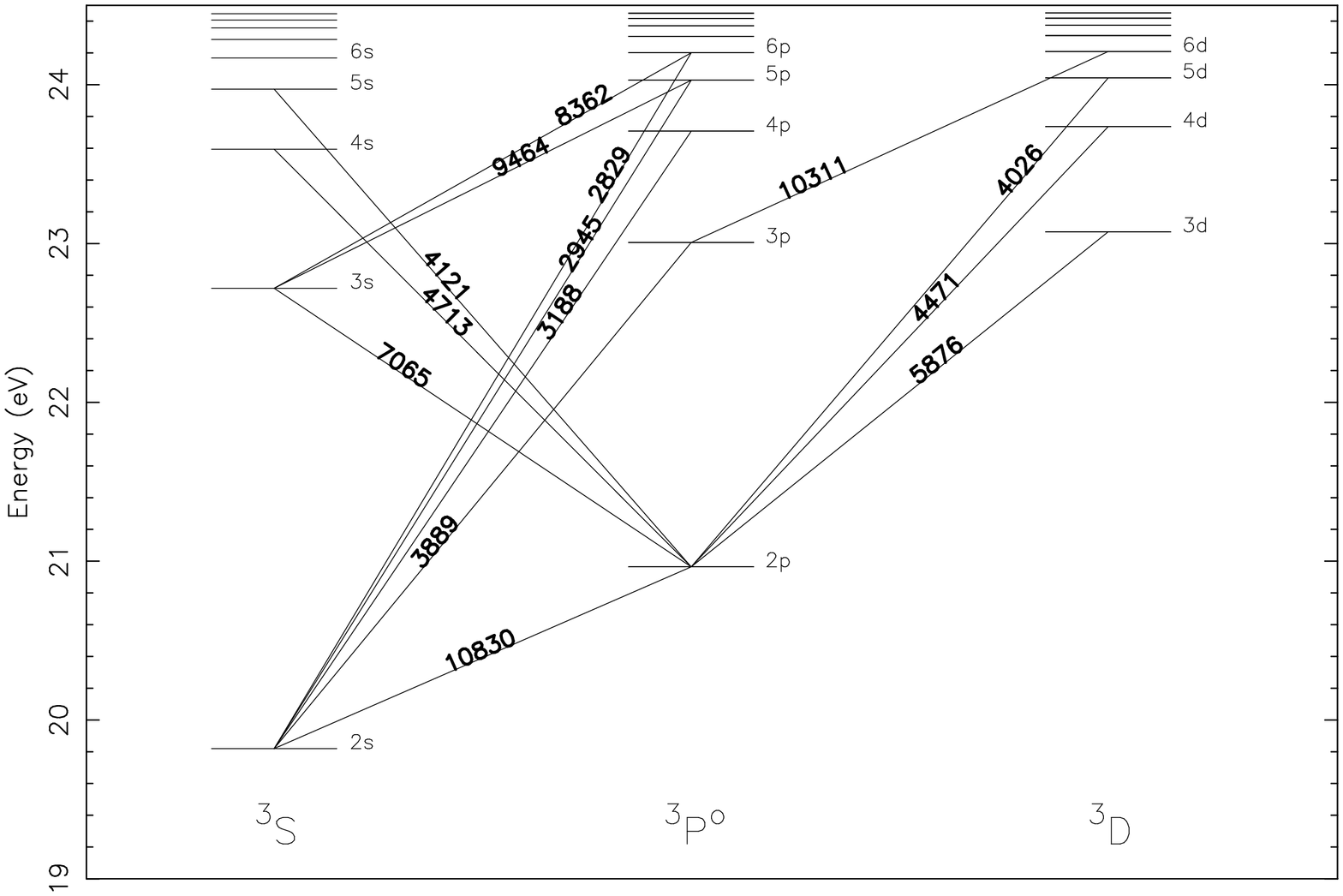}
}}
}
\caption[Grotrian diagram of He~{\sc i} singlet and triplet lines]{
Energy-level diagram of He~{\sc i}, showing $(top)$ singlet and $(bottom)$ triplet transitions.  For the sake of
clarity, not all observed transitions are shown.
\label{hei_grotrian}}
\end{figure}

Based on predictions from recombination theory
we have used the H~{\sc i} Balmer and Paschen series to calibrate the reddening 
curve.
The same
should be possible using recombination theory \citep{smi96, ben99, por05}
and observations of the He~{\sc i} lines (see energy-level diagrams in Fig.~\ref{hei_grotrian}).

\subsection{Extinction-corrected data \label{ext_corrected_data}}

Table~\ref{all_helium_table} presents our He~{\sc i} 
data from 
FOS-1SW and (the section of our STIS spectrum that covers 1SW) STIS-SLIT1c observations as 
well as complementary data from comprehensive ground-based studies by
OTV92, EPTE98, BVV00, and EPG04.

Each of the He~{\sc i} lines
has been corrected for reddening using our best-fit
curve
as normalized using the observed
H~{\sc i} Balmer (and Paschen, where available) series lines,
and each line's case~B prediction.
Case~B predictions depend on $T_e$ and $N_e$ \citep{sto95} and so
are unique to each observation set:
$T_e=9000$~K, $N_e=4000$~cm$^{-3}$ (OTV92);
$T_e=9500$~K, $N_e=5700$~cm$^{-3}$ (EPTE98, position 2);
$T_e=9000$~K, $N_e=5000$~cm$^{-3}$ (BVV00);
$T_e=8300$~K, $N_e=8900$~cm$^{-3}$ (EPG04);
$T_e=8000$~K, $N_e=2500$~cm$^{-3}$ (FOS-1SW, STIS-SLIT1c).

As is common, the line strengths are given relative to the singlet line $\lambda$4471.

The quoted uncertainties for OTV92 and EPTE98 are estimated from their
description of the quality of the data.
Those of EPG04 are directly from the per cent errors in their Table~2.
BVV00 and FOS-1SW uncertainties were determined using the 
signal to noise ratio, $S/N$,
as found from fitting each emission line to a five-parameter
(area, central wavelength, FWHM, continuum baseline and continuum slope)
single Gaussian model.  This $S/N$ is combined with the integrated line
flux to produce the uncertainty.
  The STIS data is similarly fit with a model which takes 
into account the $0\farcs5$ STIS slit width \citep[see Appendix A in][]{rub03}.  Again, the resultant
$S/N$ from this fit is combined with the integrated flux to determine the uncertainty.

\subsection{Case~B \label{caseb_section}}
He~{\sc i} case~B predictions for FOS-1SW and STIS-SLIT1c $T_e$ and $N_e$ ($n\le5$, Porter et al.\ 
2005;
$n>5$, Smits 1996)
are shown in column~(3) of Table~\ref{all_helium_table}
and are quite representative for the range of
nebular densities and temperatures we consider.

Most of the reddening-corrected singlet and triplet lines (cols.~(6)-(11))
are well-approximated by case~B recombination alone (col.~(3)).

In comparing their case~B results with the OTV92
observations, \citet{ben99} noted (their Figs.~2a and 2b)  that the lines
(singlets and triplets) shortward of $\lambda$3820 were consistently
brighter than predicted, increasingly with decreasing wavelength,
except $\lambda$3188 which appeared to be accurately predicted.
They suggested that this upward trend shortward of $\lambda$3820
could be due to unaccounted for
radiative transfer effects, though this seems very unlikely given the     
lines involved.
Inclusion of radiative transfer (see discussion below) would
cause the plotted $\lambda$3188 ratios to increase --
possibly enough to
agree with the apparent upward trend shortward
of $\lambda$3820 --
while not affecting the other lines shortward of $\lambda$3820.
If alternatively this upward trend arose from inadequate/unmodeled
extinction corrections, there would have to be a very sudden (relative)
decrease in the extinction in order for the extinction-corrected lines not
to appear so bright.
It is informative to note that the discrepancy in this instance cannot be explained
by the inaccurate extrapolation of CP70 as discussed in \S~\ref{stell_nebUV} (since CP70
is not used in the extinction correction).
   It seems most probable that there is a systematic
problem with the absolute calibration used and/or significant measurement
errors due to strong atmospheric extinction varying rapidly with
wavelength for $\lambda < 3500$~\AA\ (OTV92).  See also \S~\ref{compobshei},
Figure~\ref{heidecrementfigure}.

In their discussion of nebular model predictions of the He~{\sc i} lines compared
to the EPG04 observations, \citet{por06} have noted
that He~{\sc i} lines of the $2~^1S~-~n~^1P$ series in the same wavelength range
($\lambda\lambda 3448, 3355, 3297$) are also brighter than the expected
case~B predictions.  Analyzing the line ratios between successive members
of this series (their Figs.~1 and 5) they conclude that there are problems
with these observational results, possibly with the extinction
(over-)correction.  The same observations, corrected according to our
best-fit extinction curve, are presented in Table~\ref{all_helium_table}, where it can be seen
that the observations and case~B (or our full models) are in agreement
within the errors.

As seen in Table~\ref{all_helium_table},
use of our extinction curve brings $\lambda$7281 into agreement with the case~B  
predictions \citep[as suggested by][]{por06}.
Many of the other near-IR lines
observed by EPG04 are still
scattered about the best-fit expectation (Table~\ref{all_helium_table} and graphically
in \S~\ref{valid_IR}, Figure~\ref{est_ccm}).
These infrared lines are difficult to observe because of atmospheric
effects (OTV92) and the errors might be underestimated \citep{por06}, but
the important thing to note is that there does not appear to be any systematic discrepancy -- except
as noted below for the infrared lines of the triplet series $3~^3S~-~n~^3P$.

\subsection{Anomalous decrements \label{anom_decrements}}

Despite the validity of case~B for the prediction of most lines,
there remains a systematic mismatch between
observed and predicted ratios
for a number of the He~{\sc i} lines from
the series $2~^3S~-~n~^3P$, $2~^3P~-~n~^3S$, $3~^3S~-~n~^3P$, and $2~^1S~-~n~^1P$.

For the triplet lines, the underlying
reason is the metastability of the $2~^3S$ term which can only be depopulated
via photoionization, through collisional transitions to $2~^1S$ and $2~^1P$, or through
the (strongly forbidden) radiative transition $1~^1S~-~2~^3S$
\citep{ost89}. 
The visible/near-UV lines of the $2~^3S~-~n~^3P$ series (including $\lambda\lambda$3889, 3188, 2945, 
2829, etc.) are affected by self-absorption from the
metastable $2~^3S$ term to the corresponding $n~^3P$ term.  This results in
case~B recombination theory over-predicting these lines (see Table~\ref{all_helium_table}).
Note that although $\lambda 3889$ is blended with a Balmer line, it is
still possible to predict the observed intensity of the He~{\sc i} component and use it as a
valuable diagnostic.  For the entries in Table~\ref{all_helium_table}
we used case~B H~{\sc i} predictions and the He$^+$/H$^+$ 
derived from the other lines to deblend $\lambda$3889; this also enables a ``case~B'' prediction for
the blend (see col.~(3) in Table~\ref{all_helium_table}).  The 
space-based observations that
we report here are particularly valuable for verifying the expected
diminution of this self-absorption effect in higher members of the series.  This appears
to be the case as discussed in the modeling below.

For self-consistency, there must also be a detectable
resonance
fluorescence enhancement of the $2~^3P~-~n~^3S$ series as some of the 
electrons promoted by self-absorption 
to $n~^3P$ radiatively cascade back down to $2~^3S$ via alternate routes.
Indeed, this is quite evident in the $2~^3P~-~n~^3S$ series, primarily in $\lambda$7065 ($n=3$),
but detectable in
$\lambda$4713 ($n=4$) and $\lambda$4121 ($n=5$) too (see Table~\ref{all_helium_table}).
Similarly, we expect to see an enhancement of other triplet series including the near-IR series,
$3~^3S~-~n~^3P$.  The data in Table~\ref{all_helium_table} show this enhancement.
We will return to analysis of these series in \S~\ref{compobshei}.

The singlet lines of the series $2~^1S~-~n^1P$, $\lambda$5016 ($n=3$) (and to a much lesser 
extent $\lambda$3965, $n=4$) are also observed
to be consistently less than the case~B prediction.  As discussed in \citet{por06}, 
this is most probably a result of UV lines ($\lambda$537, $1~^1S~-~3~^1P$ and
$\lambda522$, $1~^1S~-~4~^1P$) 
escaping the nebula,
resulting in case~B being shifted slightly toward case~A \citep[as defined in][]{bak38}.

\subsection{Photoionization models \label{cloudy_extmodels} }
The CLOUDY photoionization code \citep{fer98}
accounts for many radiative transfer effects, including those relating to the
metastable
$2~^3S$ term.  It is also possible to run CLOUDY models that exclude all line transfer in order to 
investigate its importance.
With this in mind, a series of CLOUDY photoionization models was 
developed.

Self-absorption effects depend on the line width because this affects the opacity.
The He~{\sc i} lines (among others)
observed in the ground-based echelle spectra
\citep[BVV00;][]{bla06a,bla06b}
have line widths (FWHM) in excess of the Doppler/thermal and instrumental widths.
When the Doppler/thermal and instrumental widths are subtracted (in quadrature) 
from the measured FWHM 
there remains an unaccounted for contribution to the broadening which has been labeled as 
``turbulence''.
In the case of the He~{\sc i} singlet lines (which do not have any fine-structure levels
and therefore will not have any further FWHM contribution from a line blend) in the 1SW 
ground-based echelle spectra \citep{bla06b}, we find FWHM$_{turb} = 15.5\pm2.4$~km~s$^{-1}$.
This is quite consistent 
with the
``turbulence'' contribution quoted in \citet{opp03} for the He~{\sc i}~$\lambda$5876 triplet line: 
$18.4\pm2.9$~km~s$^{-1}$.  This latter line and other triplet lines have another broadening contribution 
brought on by
transitions in the fine structure ($J$ levels) and therefore are slightly less reliable in the 
determination of
the additional ``turbulence'' contribution.
As was
mentioned in \citet{opp03} and is also found here, the magnitude of the ``turbulence'' contribution to the line
width appears
to decrease as one probes along the line of sight from cooler (associated with H~{\sc i} and He~{\sc i}) to
hotter (associated with [O~{\sc iii}], [O~{\sc ii}], [S~{\sc ii}], etc.) regions of the nebula.  As the
He~{\sc i} recombination lines preferentially probe the cooler region of the nebula, the ``turbulence'' that
these lines reveal is an upper limit for the nebular model.

\begin{figure}
\epsscale{1.0}
\centerline{
\rotatebox{270}{
\scalebox{1.0}[1.0]{
\plottwo{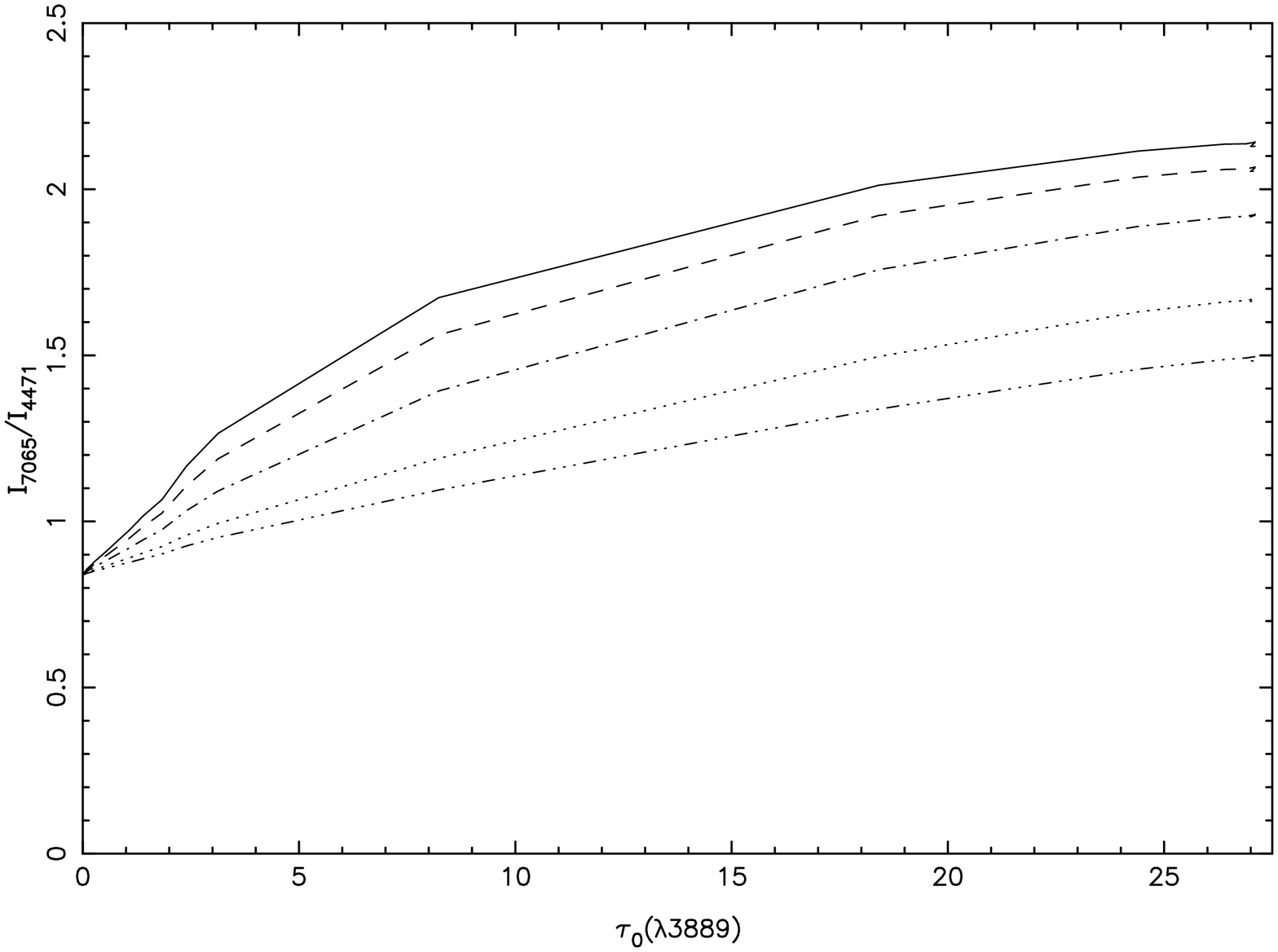}{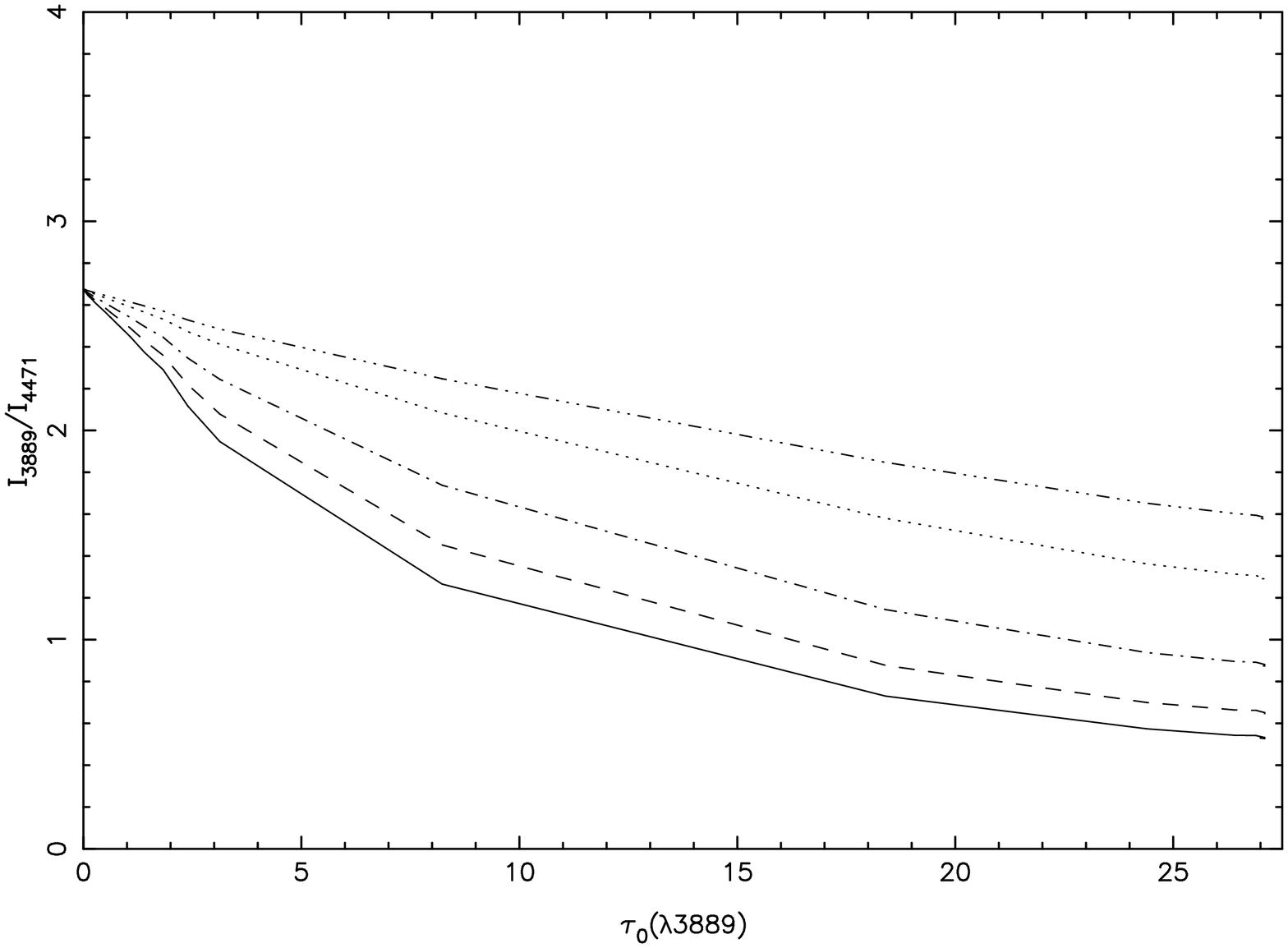}
}}
}
\caption[Variation in He~{\sc i} lines as a function of 
$\tau_0$($\lambda3889$) for different amounts of turbulence]{
Variation in line flux ($I_{\lambda}/I_{4471}$) as a function of $\tau_0$($\lambda3889$),
the optical depth at the center of $\lambda3889$ for $v_{turb}$ = 0 (solid line).
$(top)$ $I_{7065}/I_{4471}$; $(bottom)$ $I_{3889}/I_{4471}$.
For the same physical models, the changes when $v_{turb}$ = 5, 10, 20, and 
30~km~s$^{-1}$ are also shown.
\label{7065and3889tau}}
\end{figure}

This turbulent contribution to the line width
will affect the 
magnitude of the radiative transfer effect on
the $2~^3S-n~^3P$ series of lines (and the related lines, like $\lambda$7065),
as it
broadens the lines and lowers the effective optical depth and line trapping.
The effects from radiative transfer
should diminish as the turbulence is increased.
To investigate the effects of this change in
turbulence on line predictions, we developed a series of constant density ($10^4$~cm$^{-3}$)
and constant temperature ($10^4$~K)
CLOUDY models each with a different turbulence parameter, 
$v_{turb}=\mathrm{FWHM}_{turb}/\sqrt{4 \ln 2}$ (as defined in CLOUDY), and excluding ``induced processes''
(as defined in CLOUDY, these include continuum fluorescent excitation and induced recombination).
Then each of these constant-turbulence 
models was varied in thickness,
in order to develop plots of line ratios as a function of 
$\tau_0$($\lambda3889$) -- the optical depth at the center of He~{\sc i}~$\lambda$3889 for zero turbulence 
velocity
(see Fig.~\ref{7065and3889tau}).
For the normalizing line we use $\lambda$4471 which is not significantly affected by radiative transfer.
Our Figure~\ref{7065and3889tau} can be compared 
with Figure 4.5 of \citet{ost89}, which shows similar variation, but in the latter case as a
result of changes in the nebular expansion velocity.
At $\tau_0=0$ all models reduce, as is expected, to their case~B values. 

\begin{figure}
\epsscale{0.7}  
\centerline{
\rotatebox{270}{
\scalebox{1.0}[1.0]{
\plotone{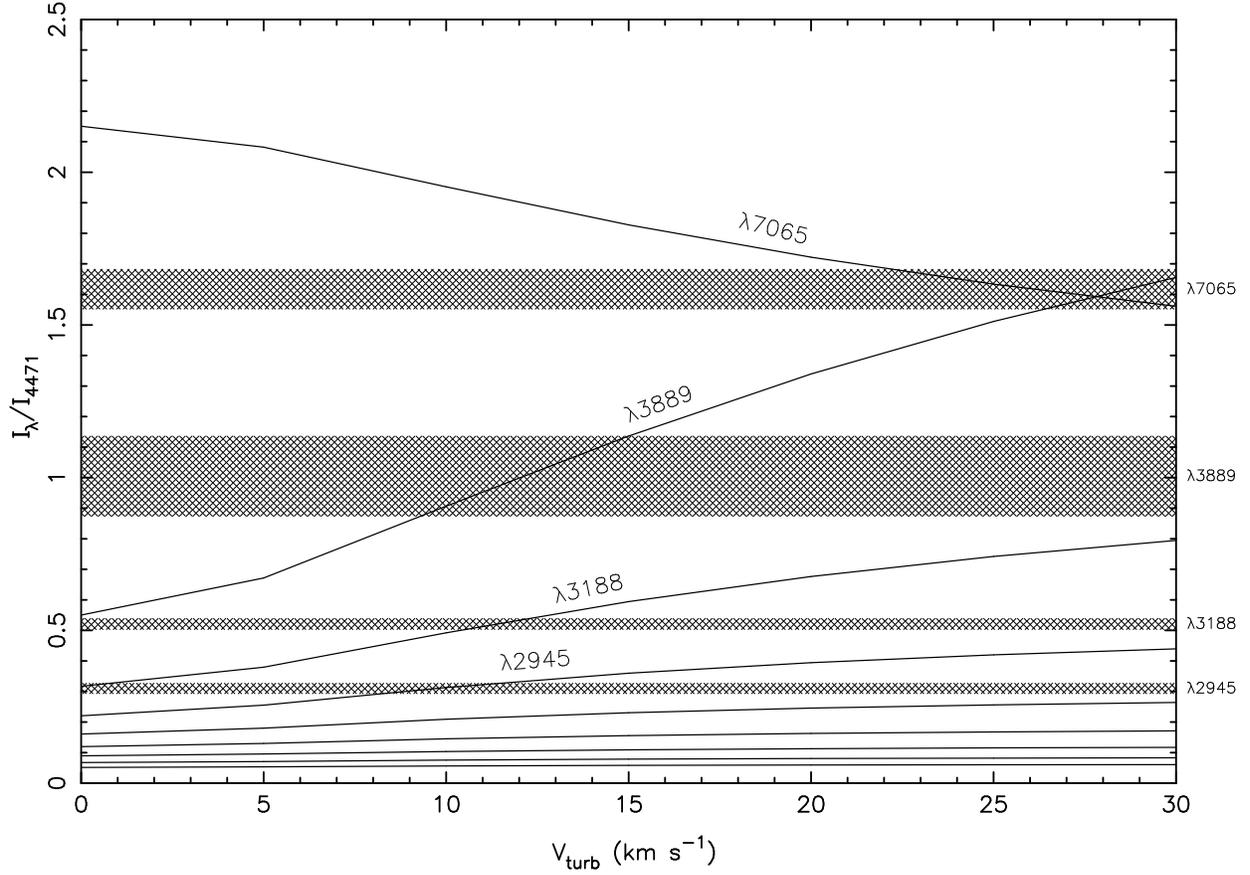}
}}
}
\caption[Variation of He~{\sc i} lines as a
function of 
$v_{turb}$ in the full nebular model]{Variation of He~{\sc i} lines ($2~^3S~-~n~^3P$ series) as a 
function of
$v_{turb}$ in the full nebular model.
The hatched boxes represent the dereddened FOS-1SW observations (and uncertainties)
of $\lambda\lambda7065, 3889$ (deblended), $3188$, and $2945$ (from Table~\ref{all_helium_table}).
\label{allheitau}}
\end{figure}

Figure~\ref{allheitau} 
plots the line strengths, $I_{\lambda}/I_{\lambda4471}$,
as a function of each 
model's ``turbulence'' parameter
for fixed $\tau_0$($\lambda3889$)$\sim27$ which is appropriate to the column density of our full nebular model 
below.  In this case
to enable the
best comparison between
the CLOUDY model predictions and our observations,
the ``induced processes'' are included
(regardless, in our full model the relative importance of the induced processes is much diminished because
 the nebula becomes optically thick to the continuum radiation responsible for these 
excitations).
From the overplotted observed $I_{\lambda7065}/I_{\lambda4471}$ in Figure~\ref{allheitau},
we would predict $v_{turb}=20-30$~km~s$^{-1}$, whereas from $\lambda3889$ (deblended), $\lambda3188$ and $\lambda2945$,
$v_{turb}=10-12$~km~s$^{-1}$.  This latter prediction is consistent with the ``turbulence'' 
contribution (FWHM$_{turb}\sim16$~km~s$^{-1}$; i.e., $v_{turb}\sim10$~km~s$^{-1}$)
which was determined from the observed 
He~{\sc i} lines' FWHM above.  The value $v_{turb}=12$~km~s$^{-1}$ will be adopted in the CLOUDY 
models discussed hereafter.  Models with this value of the turbulence should be able to roughly explain the 
observations of the
hitherto anomalous
He~{\sc i} lines.
As expected, the results of the full nebular models to be discussed (Table~\ref{all_helium_table})
are slightly different than the results for constant $N_e$ and $T_e$ in Figure~\ref{allheitau},
in particular giving even better agreement for $\lambda$7065.

Our full nebular models are similar to the Orion Nebula model
discussed in \citet{bal91} -- i.e., a closed geometry and constant pressure.
Two such models (see Tables~\ref{cloudy_hei} and \ref{abund_table_hei})
were created, as was done
in \citet{bla06a}: one with a Mihalas stellar atmosphere model \citep{mih72} and H~{\sc ii} 
region abundances as defined in CLOUDY
\citep[][OTV92]{bal91,rub91} (model~M); and one with a Kurucz 
stellar 
atmosphere model \citep{kur79} and 
EPG04 Orion Nebula abundances (model~K).
The Mihalas (non-LTE) stellar atmosphere has been shown to best represent the incident 
continuum radiation 
(BVV00), but to test the robustness of our results we also include the 
Kurucz (LTE, line-blanketed) 
stellar atmosphere in our second model.
Each of these models 
has $v_{turb}=12$~km~s$^{-1}$.
The other parameters (density, radius, etc.) were varied so as to best reproduce the
SLIT1c
\hbeta\ surface brightness
and the temperature- and density-sensitive lines, as 
summarized in
Table~\ref{constraint_hei} for both models.

The models' predictions of the He~{\sc i} lines given in columns~(4) and (5) in 
Table~\ref{all_helium_table}
include the deviation
from case~B introduced by the metastable $2~^3S$ term.  The effect of this metastability
is reflected in the models'
(non-zero) line 
optical depth, $\tau$($\lambda3889$)=13.
Note that He~{\sc i} $\lambda$3889 ($n=3$) appears as a blend with H~{\sc i}
$\lambda$3889 and He~{\sc i}
$\lambda$2829 ($n=6$) appears as a blend with [Fe~{\sc iv}] $\lambda$2829.
Neither of these blends is included in our quantitative analysis, but the predicted value for each is
 shown in a separate row in 
Table~\ref{all_helium_table}.

Models~M and K were modified to exclude radiative transfer in order to
further investigate the validity of
the CLOUDY models.  
This exclusion
has the expected effect on the lines: returning the triplet line 
predictions to roughly their 
case~B values and those singlets connected radiatively to the ground state are reduced to their
case~A values.

\subsection{Comparison with observations \label{compobshei}}
We have modeled the extinction correction for OTV92, EPTE98, BVV00, EPG04 and our FOS-1SW
and STIS-SLIT1c observations using our extinction curve and
the H~{\sc i} Balmer (and Paschen, where available) lines and
a subset of He~{\sc i} lines,
excluding lines from the triplet series $2~^3S~-~n~^3P$ ($\lambda\lambda$3889, 3188, 2945, 2829), $2~^3P~-~n~^3S$ 
($\lambda\lambda$7065, 4713, 4121) and $3~^3S~-~n~^3P$ (all lines) and the singlet line, $\lambda$5016.
Our $C_{\mathrm{H}\beta}$ from the fit of H~{\sc i} and the subset of He~{\sc i} lines
is consistent with that found using H~{\sc i} lines alone (see end of Table~\ref{all_helium_table}),
suggesting
that both H~{\sc i} and this subset of He~{\sc i} lines are reliable
determinants of the extinction correction. 

\begin{figure}
\epsscale{0.8}
\centerline{
\rotatebox{270}{
\scalebox{1.0}[1.0]{
\plotone{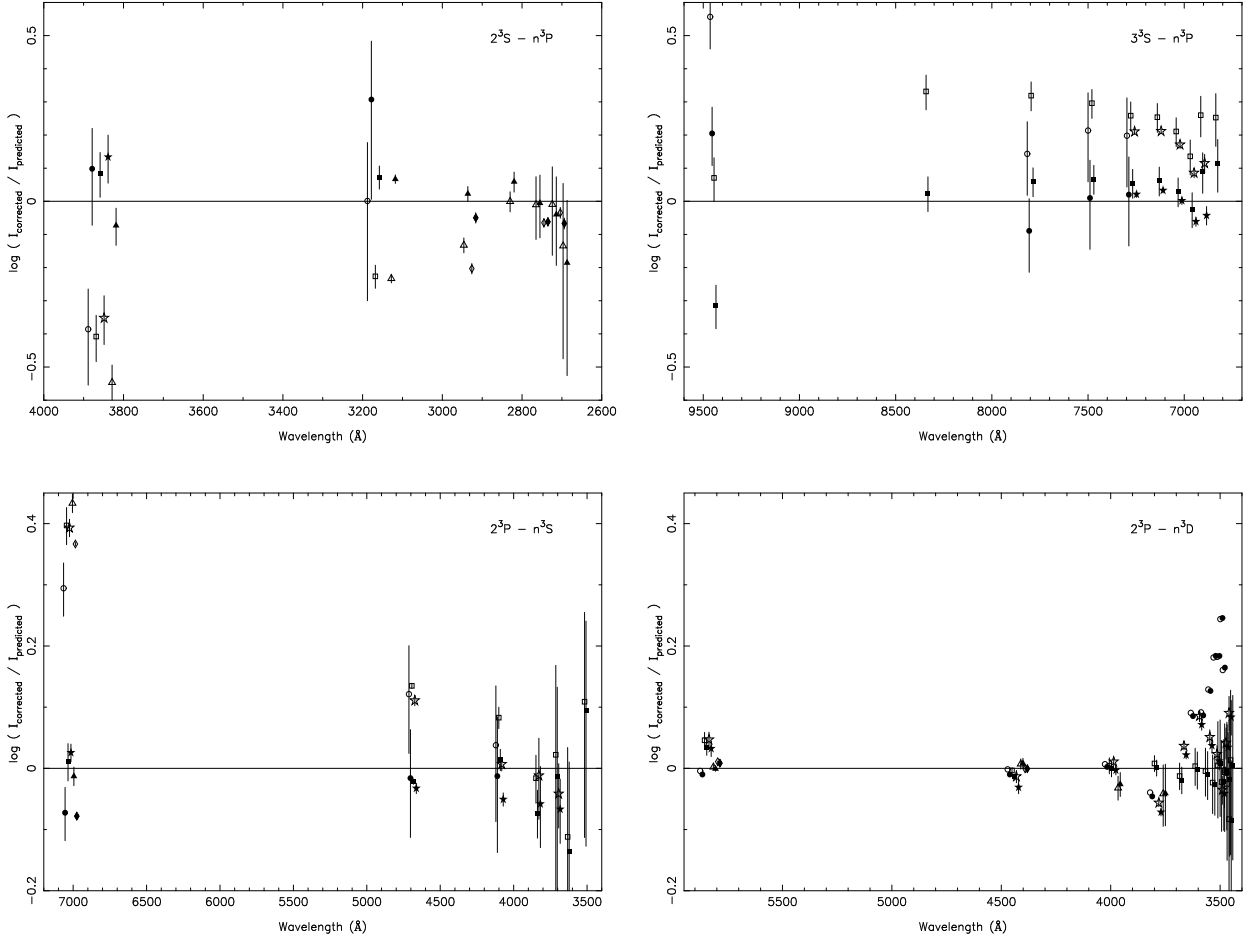}
}}
}
\caption[Observed He~{\sc i} decrements compared to case~B and CLOUDY model predictions for four series]{Observed
He~{\sc i} 
decrements (OTV92, $\bullet$; BVV00, $\star$; EPG04, $\blacksquare$; FOS-1SW, $\blacktriangle$; STIS-SLIT1c, $\blacklozenge$)
compared to case~B (open symbols) and CLOUDY model~M (filled symbols) 
predictions for four series.  The corrected values, $I_{corrected}$, are
determined from our extinction curve, which is calibrated for each dataset using H~{\sc i} and a subset of He~{\sc i} lines (see
\S~\ref{compobshei}).  The 68.3\% confidence intervals are shown for each member, $n$, of the series.
For visual clarity, each set of observed data has been offset by 10\AA, and
the uncertainties have been excluded for the 
OTV92 data in the $2~^3P~-~n~^3D$ series (in the UV, these uncertainties are close to
$\pm0.2$~dex).
Note the change in scale between the top and bottom figures.
\label{heidecrementfigure}
}
\end{figure}

\begin{figure}
\epsscale{0.7} 
\centerline{  
\rotatebox{270}{
\scalebox{1.0}[1.0]{
\plotone{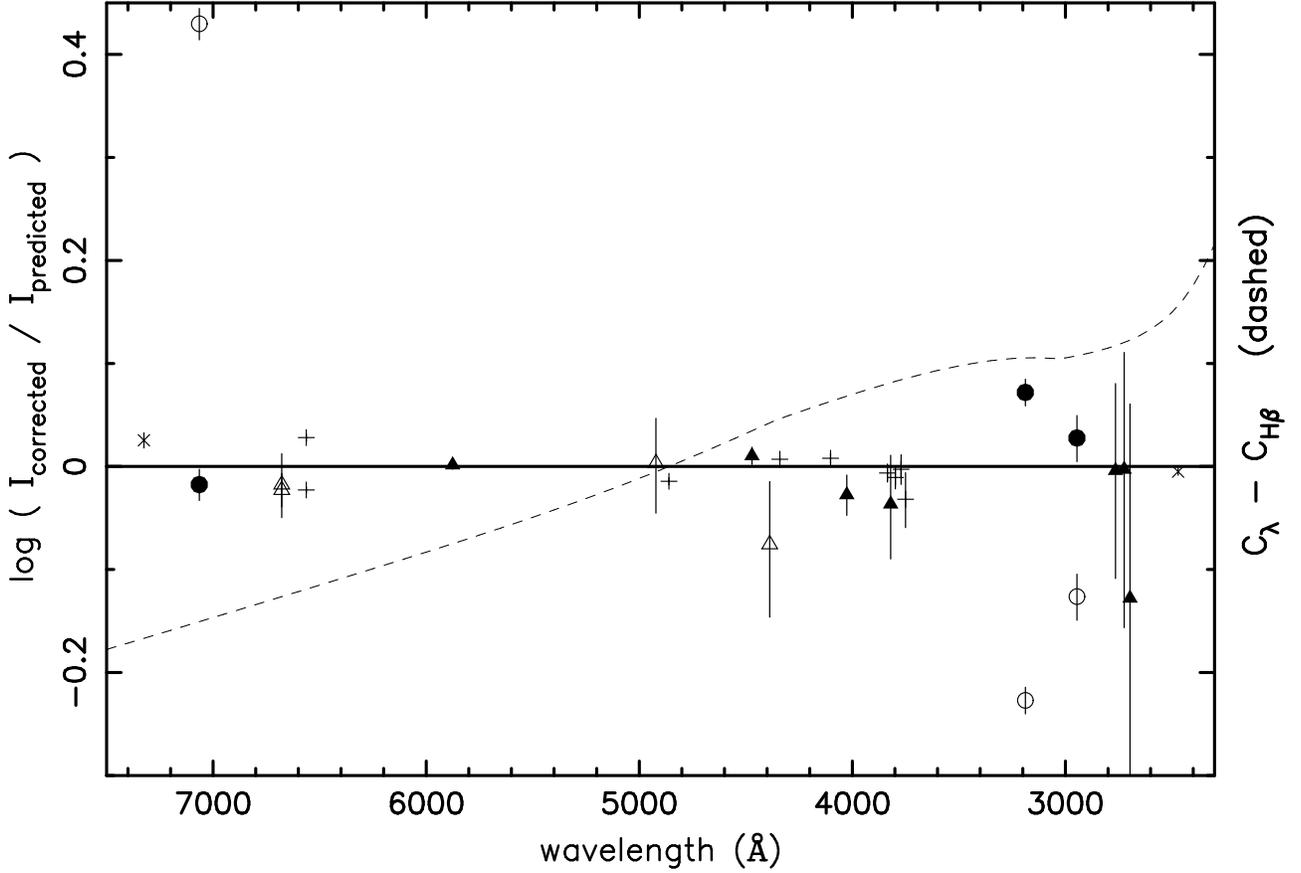}
}}
}
\caption[Differential extinction curve for FOS-1SW]{Differential extinction curve for FOS-1SW (dashed line) presented as 
$C_{\lambda}-C_{\mathrm{H}\beta}$.
Residuals, log($I_{corrected}/I_{predicted}$) (relative to case~B), and $1\sigma$ errors
 are plotted for each emission line used in the fit:  H~{\sc i} (+), He~{\sc i} (triplets, $\blacktriangle$; singlets, $\triangle$)
and [O~{\sc ii}] ($\times$).
The three He~{\sc i} lines most affected by radiative transfer ($\lambda\lambda$7065, 3188, and 2945) are not 
included in the fit, but are 
plotted relative to case~B values ($\circ$) and model~M predictions ($\bullet$).
\label{FOSfit} }
\end{figure}

\begin{figure}
\epsscale{0.7}
\centerline{
\rotatebox{270}{
\scalebox{1.0}[1.0]{
\plotone{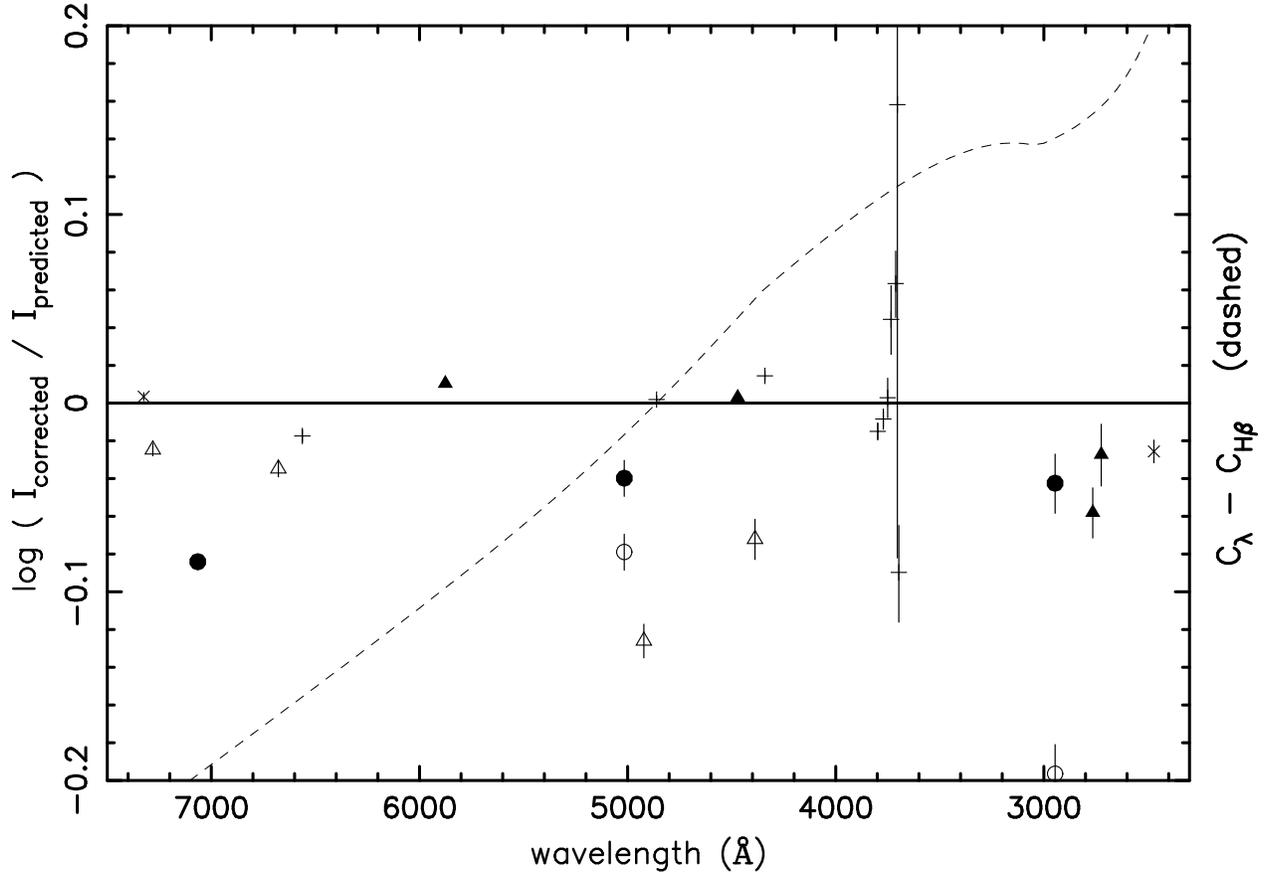}
}}
}
\caption[Differential extinction curve for STIS-SLIT1c]{Like Fig.~\ref{FOSfit}, but for
STIS-SLIT1c and a different scale.
The two He~{\sc i} lines most affected by radiative transfer ($\lambda\lambda$7065 and 2945) are not
included in the fit, but are
plotted relative to case~B values ($\circ$) and model~M predictions ($\bullet$).
For case~B, $\lambda$7065 has log($I_{corrected}/I_{predicted}$)$=0.4$.
Similarly  shown is singlet $\lambda$5016 which was also not included in the fit.
\label{STISfit} }
\end{figure}

For comparison with these extinction-corrected observations, $I_{corrected}$, 
we adopt separately predictions, $I_{predicted}$, from case~B and CLOUDY model~M.
Values of log($I_{corrected}/I_{predicted}$)
are plotted in
Figure~\ref{heidecrementfigure} for all observed members of four
series for both case~B and model~M predictions.
Many data sets are shown, but EPTE98 is excluded simply to avoid visual clutter.
For the $2~^3P~-~n~^3D$ series, which includes $\lambda$4471 ($n=4$), case~B and model~M are in close agreement.
This triplet series is not significantly affected by radiative transfer effects and the theory and observations are in close 
agreement.  The other triplet series are clearly affected, by self-absorption ($2~^3S~-~n~^3P$) or by resonance
fluorescence enhancement ($2~^3P~-~n~^3S$, $3~^3S~-~n~^3P$).
Case~B is a poor approximation, but the agreement between theory and observations
improves dramatically as we switch from case~B to model~M.  ($\chi^2$ is reduced by a factor of 
roughly 10.)

Two features deserve further discussion.  The OTV92 data shortward of 3700~\AA\ (including $\lambda$3188) 
most likely suffer from systematic errors (\S~\ref{caseb_section}).  This can be seen graphically 
in the $2~^3P~-~n~^3D$ series of Figure~\ref{heidecrementfigure}
when
comparing OTV92 (circles, no uncertainty bars) and others' datasets; the OTV92 data points (with one exception, 
$\lambda$3479) all lie 
significantly apart from the other data.
Second, $\lambda$9464 measured by EPG04 appears to be too faint.  This line has
a number of neighboring near-IR H~{\sc i} Paschen lines and He~{\sc i} lines
which appear to be well-fit by our extinction correction.
There would have to be a large sudden increase in the amount of extinction for
the extinction correction to be   
 able to account for this anomaly.
This suggests that $\lambda$9464 may have a larger than reported measurement
uncertainty \citep[see also \S~\ref{caseb_section} and][]{por06}.

\subsection{Extinction corrections revisited \label{ext_corr_revisited}}

The STIS and FOS data are corrected for extinction using our extinction
curve, the unblended H~{\sc i} lines, the subset of He~{\sc i} lines and the common-upper-level pair
of [O~{\sc ii}] lines (refer to \S~\ref{compobshei} and \S~\ref{validUV}).
These are presented in Table~\ref{corrected_table} relative to the H$\beta$ predicted from the fit,
along with the derived $C_{\mathrm{H}\beta}$.

In Figure~\ref{FOSfit},
we show the extinction-corrected FOS-1SW line intensities compared to
the case~B theory (the residuals of the fit) as well as the extinction
correction applied (here expressed differentially with respect to
$C_{\mathrm{H}\beta}$). 
The He~{\sc i} lines $\lambda\lambda$7065, 3188, and 2945 are excluded
from the fit, but their values relative to case~B 
and model~M predictions are overplotted.  Model~M provides a better
prediction for the
lines affected by radiative transfer.

The analysis for the 
SLIT1c data is shown in a similar way in Figure~\ref{STISfit}.
The He~{\sc i} lines $\lambda\lambda$7065, 5016 and 2945 (and H~{\sc i} line $\lambda$3704,
because of large uncertainty) are excluded
from the fit, but their values relative to case~B and model~M
 are overplotted. 
 Again, model~M provides a better
prediction for the      
lines affected by radiative transfer.

In fitting the UV extinction curve, we conclude that
$\lambda$3889 (a blend in any case), $\lambda$3188, $\lambda$2945, and
$\lambda$2829 (also a blend) should be
excluded -- pending
correction for radiative transfer effects -- but that lines associated with higher $n$
(i.e., $n>6$) can be
included to ensure a more robust determination.
The visible/IR extinction correction can be strengthened by including case~B predictions for
He~{\sc i} triplet and singlet
lines, but as discussed above, we conclude that lines from the series $3~^3S~-~n~^3P$ (all lines), $2~^3P~-~n~^3S$
($\lambda\lambda$7065, 4713, 4121) and $2~^1P~-~n~^1S$ ($\lambda$5016) should be excluded
-- again pending correction for radiative transfer effects.

\subsection{He$^{+}$/H$^{+}$ ratio \label{heh_ratio}}
In fitting the H~{\sc i} and He~{\sc i} lines simultaneously, we are also able to calculate the He$^{+}$/H$^{+}$ abundance 
directly
from the parameterization used in the fitting
(see \S~\ref{parameterization}).
These abundances are included at the
bottom of Tables~\ref{all_helium_table} and \ref{corrected_table}.
Table~\ref{all_helium_table} also shows
the abundances from the original reference (when published).
Our new analysis agrees with these within the errors.  Our confidence intervals
reflect implicitly the consistency of the extinction correction as well as
the match to theory and observational errors and so are somewhat larger.

The He$^{+}$/H$^{+}$ ratio could in principle vary with position in the nebula if the ionization structure
(hence the ionization correction factor, ICF, used to convert He$^{+}$/H$^{+}$ to He/H) differs on account
of nebular structure.  However, this is likely to be a small effect.  It is the case that the He$^{+}$/H$^{+}$
values for the different positions agree with one another within their errors.
The unweighted mean and its standard deviation is $0.0883\pm0.0019$.  If weighted, the ratio is
$0.0882\pm0.0004$.
Neither accounts for any systematic error.

Extensive discussion of the ICF required to determine the atomic He/H ratio 
would take us beyond the scope of this paper.
However, our CLOUDY model~M gives 1.14 and model~K gives 1.00, which are within the range given by other authors;
1.1 is a typical value with an uncertainty of roughly $\pm0.05$.  
Adopting this we obtain He/H = $0.097\pm0.002\pm0.005$.
 The latter uncertainty from the ICF is probably the major source of uncertainty,
greater than that in measuring the ionic abundance.

\section{Summary}
Using slight modifications to the valuable stellar extinction curve developed by CCM89, we have
been able to develop an accurate new analytic method of determining the nebular extinction curve
over an extensive wavelength range in the Orion Nebula.
This curve has been rigorously
tested with currently available near-IR, optical, and
ultraviolet ground-based and space-based data,
standing up as a robust measure of the extinction.
We have also compared this new curve with the
CP70 nebular extinction curve.
We have confirmed that the discrepancy, with respect to theoretical expectations,
in some near-IR
and near-UV He~{\sc i} lines measured by EPG04 is
a result of
inaccurate extinction correction,
but that the UV discrepancy in observations by OTV92
is a result of calibration and/or measurement uncertainty and not an extinction (or radiative transfer) effect. 

On the foundation of this new extinction analysis,
we have measured systematic anomalous He~{\sc i} decrements, compared to case~B, associated
with the $2~^3S-n~^3P$, $2~^3P-n~^3S$ and $3~^3S-n~^3P$ series, and to a lesser extent with the $2~^1S-n~^1P$ series,
none of which can be explained by any adjustment to the extinction curve.
Qualitatively, these anomalies are as expected from radiative transfer effects, for the triplets arising
from the metastability of $2^3S$ \citep{ost89}.
Furthermore, modeling of the radiative transfer effects using CLOUDY produces a remarkable
quantitative agreement between theory and observation.

Because He~{\sc i} case~B recombination theory is not reliable  for a subset of He~{\sc i}
permitted lines associated with these aforementioned series,
those lines most affected ($\lambda$3889,
$\lambda$3188, $\lambda$2945, $\lambda$2829, $\lambda$7065,
$\lambda$4713, $\lambda$4121, $\lambda$5016 and those of the triplet 
series $3~^3S~-~n~^3P$) 
must be excluded in the
determination of the amplitude of the extinction curve,
or in the
calculation of He$^+$/H$^+$ abundance.
 Alternatively, they could be included after
 adjustment from their case~B predictions by modeling radiative transfer effects.

\acknowledgments

This work was supported in part by the Natural Sciences and Engineering
Research Council of Canada.  We thank STScI for its support through
grants GO-4385, GO-5748, GO-6056, GO-6093, GO-7514 and AR-06403.01-95A.  RR
was partly supported by an LTSA grant.  RJD
acknowledges NASA/Ames Research Center Interchange Grants NCC2-5008.

\clearpage
\input{tab1.tex}

\clearpage
\input{tab2.tex}

\clearpage
\input{tab3.tex}

\clearpage
\input{tab4.tex}

\clearpage
\input{tab5.tex}

\clearpage
\input{tab6.tex}

\clearpage
\input{tab7.tex}

\clearpage
\input{tab8.tex}

\clearpage

\clearpage  

\clearpage

\clearpage

\clearpage

\clearpage

\clearpage

\clearpage

\clearpage

\clearpage

\end{document}

%% file: tab1.tex
\begin{deluxetable}{clrrrrr}
\tabletypesize{\tiny}
\tablewidth{0pt}
\tablecolumns{7}
\tablecaption{HST STIS V2 and FOS observations prior to reddening correction ($I_{\lambda}/I_{\mathrm{H}\beta}$) \label{uncorrected_table}}
\tablehead{
\colhead{ID} & \colhead{$\lambda$ (\AA)} & 
\colhead{SLIT1b} & \colhead{SLIT1c} & 
\colhead{SLIT2b} & \colhead{SLIT2c} & \colhead{FOS-1SW}
}
\startdata
{\rm C~{\sc iii}]} & $1907$ & $  0.0371\pm0.0021$ & $  0.0534\pm0.0012$ & \nodata & \nodata & \nodata \\
{\rm C~{\sc iii}]} & $1909$ & $  0.0316\pm0.0020$ & $  0.0444\pm0.0008$ & \nodata & \nodata & \nodata \\
{\rm [O~{\sc ii}]} & $2470$ & $  0.0393\pm0.0013$ & $  0.0752\pm0.0013$ & $  0.0323\pm0.0014$ & $  0.0454\pm0.0012$ & $  0.1164\pm0.0009$ \\
{\rm He~{\sc i}} & $2697$ & \nodata & \nodata & \nodata & \nodata & $0.0017\pm0.0009  $ \\
{\rm He~{\sc i}} & $2723$ & $  0.0023\pm0.0049$ & $  0.0029\pm0.0001$ & \nodata & \nodata & $  0.0032\pm0.0010$ \\
{\rm He~{\sc i}} & $2764$ & $  0.0048\pm0.0004$ & $  0.0041\pm0.0001$ & \nodata & \nodata & $  0.0048\pm0.0010$ \\
{\rm He~{\sc i}}+{\rm [Fe~{\sc iv}]} & $2829$ & \nodata & \nodata & \nodata & \nodata & $0.0081\pm0.0006$ \\
{\rm He~{\sc i}} & $2945$ & $  0.0083\pm0.0007$ & $  0.0091\pm0.0003$ & \nodata & \nodata & $  0.0110\pm0.0006$ \\
{\rm He~{\sc i}} & $3188$ & \nodata & \nodata & \nodata & \nodata & $0.0186\pm0.0006$ \\
{\rm H~{\sc i}} & $3697$ & $  0.0114\pm0.0231$ & $  0.0068\pm0.0004$ & $  0.0127\pm0.0102$ & $  0.0044\pm0.0088$ & \nodata \\
{\rm H~{\sc i}} & $3704$ & $  0.0108\pm0.0098$ & $  0.0143\pm0.0061$ & $  0.0187\pm0.0042$ & $  0.0140\pm0.0017$ & \nodata \\
{\rm H~{\sc i}} & $3712$ & $  0.0109\pm0.0012$ & $  0.0139\pm0.0006$ & $  0.0101\pm0.0125$ & $  0.0101\pm0.0031$ & \nodata \\
{\rm H~{\sc i}}+{\rm [S~{\sc iii}]} & $3722$ & $  0.0212\pm0.0011$ & $  0.0278\pm0.0005$ & $  0.0223\pm0.0008$ & $  0.0268\pm0.0016$ & \nodata \\
{\rm [O~{\sc ii}]} & $3726$ & $  0.4797\pm0.0142$ & $  0.6041\pm0.0148$ & $  0.4898\pm0.0137$ & $  0.5228\pm0.0184$ & \nodata \\
{\rm [O~{\sc ii}]} & $3729$ & $  0.2309\pm0.0213$ & $  0.2649\pm0.0220$ & $  0.2456\pm0.0184$ & $  0.2546\pm0.0124$ & \nodata \\
{\rm H~{\sc i}} & $3734$ & $  0.0212\pm0.0019$ & $  0.0204\pm0.0009$ & $  0.0147\pm0.0005$ & $  0.0180\pm0.0012$ & \nodata \\
{\rm H~{\sc i}} & $3750$ & $  0.0190\pm0.0016$ & $  0.0236\pm0.0006$ & $  0.0243\pm0.0021$ & $  0.0190\pm0.0008$ & $  0.0242\pm0.0015$ \\
{\rm H~{\sc i}} & $3771$ & $  0.0230\pm0.0011$ & $  0.0299\pm0.0005$ & $  0.0281\pm0.0011$ & $  0.0276\pm0.0007$ & $  0.0338\pm0.0011$ \\
{\rm H~{\sc i}} & $3798$ & $  0.0308\pm0.0008$ & $  0.0395\pm0.0006$ & $  0.0297\pm0.0009$ & $  0.0361\pm0.0010$ & $  0.0444\pm0.0011$ \\
{\rm He~{\sc i}} & $3820$ & \nodata & \nodata & \nodata & \nodata & $ 0.0085\pm0.0010  $ \\
{\rm He~{\sc i}}+{\rm H~{\sc i}} & $3889$ & \nodata & \nodata & \nodata & \nodata & $0.1298\pm0.0017  $ \\
{\rm He~{\sc i}}+{\rm H~{\sc i}}+{\rm [Ne~{\sc iii}]} & $3965$ & \nodata & \nodata & \nodata & \nodata & $0.1886\pm0.0026  $ \\
{\rm He~{\sc i}} & $4026$ & \nodata & \nodata & \nodata & \nodata & $0.0167\pm0.0007  $ \\
{\rm H~{\sc i}} & $4340$ & $  0.4004\pm0.0048$ & $  0.4170\pm0.0048$ & $  0.3777\pm0.0035$ & $  0.4064\pm0.0059$ & $  0.4444\pm0.0021$ \\
{\rm [O~{\sc iii}]} & $4363$ & $  0.0091\pm0.0003$ & $  0.0095\pm0.0002$ & $  0.0074\pm0.0003$ & $  0.0115\pm0.0003$ & \nodata \\
{\rm He~{\sc i}} & $4388$ & $  0.0048\pm0.0002$ & $  0.0045\pm0.0001$ & $  0.0039\pm0.0002$ & $  0.0048\pm0.0001$ & $  0.0042\pm0.0006$ \\
{\rm He~{\sc i}} & $4471$ & $  0.0397\pm0.0005$ & $  0.0432\pm0.0005$ & $  0.0359\pm0.0005$ & $  0.0398\pm0.0006$ & $  0.0416\pm0.0008$ \\
{\rm H~{\sc i}} & $4861$ & $  1.0000\pm0.0111$ & $  1.0000\pm0.0110$ & $  1.0000\pm0.0073$ & $  1.0000\pm0.0131$ & $  1.0000\pm0.0081$ \\
{\rm He~{\sc i}} & $4922$ & $  0.0093\pm0.0010$ & $  0.0098\pm0.0002$ & $  0.0142\pm0.0010$ & $  0.0106\pm0.0007$ & $  0.0121\pm0.0013$ \\
{\rm [O~{\sc iii}]} & $4959$ & $  1.2387\pm0.0148$ & $  1.0750\pm0.0115$ & $  1.2212\pm0.0096$ & $  1.2046\pm0.0165$ & \nodata \\
{\rm [O~{\sc iii}]} & $5007$ & $  3.7521\pm0.0408$ & $  3.2426\pm0.0345$ & $  3.6822\pm0.0270$ & $  3.6453\pm0.0473$ & \nodata \\
{\rm He~{\sc i}} & $5016$ & $  0.0271\pm0.0008$ & $  0.0233\pm0.0006$ & $  0.0214\pm0.0006$ & $  0.0206\pm0.0005$ & \nodata \\
{\rm [Cl~{\sc iii}]} & $5518$ & $  0.0049\pm0.0003$ & $  0.0039\pm0.0001$ & $  0.0028\pm0.0002$ & $  0.0036\pm0.0003$ & \nodata \\
{\rm [Cl~{\sc iii}]} & $5538$ & $  0.0075\pm0.0002$ & $  0.0064\pm0.0001$ & $  0.0048\pm0.0001$ & $  0.0063\pm0.0001$ & \nodata \\
{\rm [N~{\sc ii}]} & $5755$ & $  0.0095\pm0.0003$ & $  0.0148\pm0.0002$ & $  0.0081\pm0.0003$ & $  0.0107\pm0.0002$ & \nodata \\
{\rm He~{\sc i}} & $5876$ & $  0.1797\pm0.0020$ & $  0.1703\pm0.0018$ & $  0.1643\pm0.0010$ & $  0.1812\pm0.0023$ & $  0.1449\pm0.0013$ \\
{\rm [N~{\sc ii}]} & $6548$ & $  0.2341\pm0.0031$ & $  0.2667\pm0.0031$ & $  0.2101\pm0.0021$ & $  0.2461\pm0.0035$ & \nodata \\
{\rm H~{\sc i}} & $6563$ & $  4.4094\pm0.0510$ & $  3.9699\pm0.0498$ & $  3.9989\pm0.0294$ & $  4.4652\pm0.0665$ & $  4.2067\pm0.0336$\tablenotemark{a} \\
{\rm H~{\sc i}} & $6563$ & \nodata & \nodata & \nodata & \nodata & $3.7418\pm0.0265$\tablenotemark{b} \\
{\rm [N~{\sc ii}]} & $6583$ & $  0.7209\pm0.0102$ & $  0.8343\pm0.0098$ & $  0.6460\pm0.0071$ & $  0.7654\pm0.0106$ & \nodata \\
{\rm He~{\sc i}} & $6678$ & $  0.0549\pm0.0008$ & $  0.0504\pm0.0007$ & $  0.0476\pm0.0004$ & $  0.0539\pm0.0009$ & $  0.0439\pm0.0031$\tablenotemark{a} \\
{\rm He~{\sc i}} & $6678$ & \nodata & \nodata & \nodata & \nodata & $0.0433\pm0.0016$\tablenotemark{b} \\
{\rm [S~{\sc ii}]} & $6716$ & $  0.0344\pm0.0007$ & $  0.0318\pm0.0004$ & $  0.0358\pm0.0007$ & $  0.0337\pm0.0006$ & \nodata \\
{\rm [S~{\sc ii}]} & $6731$ & $  0.0637\pm0.0012$ & $  0.0636\pm0.0008$ & $  0.0660\pm0.0009$ & $  0.0662\pm0.0011$ & \nodata \\
{\rm He~{\sc i}} & $7065$ & $  0.1176\pm0.0013$ & $  0.1058\pm0.0012$ & $  0.1036\pm0.0008$ & $  0.1169\pm0.0015$ & $  0.1014\pm0.0035$ \\
{\rm [Ar~{\sc iii}]} & $7136$ & $  0.2679\pm0.0034$ & $  0.2457\pm0.0029$ & $  0.2298\pm0.0020$ & $  0.2742\pm0.0040$ & \nodata \\
{\rm C~{\sc ii}} & $7236$ & $  0.0048\pm0.0004$ & $  0.0041\pm0.0002$ & $  0.0038\pm0.0002$ & $  0.0041\pm0.0004$ & \nodata \\
{\rm O~{\sc i}} & $7254$ & $  0.0022\pm0.0002$ & $  0.0019\pm0.0001$ & $  0.0014\pm0.0001$ & $  0.0022\pm0.0001$ & \nodata \\
{\rm He~{\sc i}} & $7281$ & $  0.0119\pm0.0003$ & $  0.0112\pm0.0001$ & $  0.0095\pm0.0002$ & $  0.0113\pm0.0002$ & \nodata \\
{\rm [O~{\sc ii}]} & $7319$ & $  0.1144\pm0.0014$ & $  0.1547\pm0.0019$ & $  0.0896\pm0.0010$ & $  0.1215\pm0.0020$ & $  0.3344\pm0.0056$ \\
{\rm [O~{\sc ii}]} & $7330$ & $  0.0936\pm0.0012$ & $  0.1276\pm0.0015$ & $  0.0723\pm0.0008$ & $  0.1006\pm0.0016$ & Blend \\
\enddata
\tablenotetext{a}{FOS/G780 grating}
\tablenotetext{b}{FOS/G570 grating}
\end{deluxetable}

%% file: tab2.tex
\begin{deluxetable}{cclr}
\tabletypesize{\tiny}
\tablewidth{0pt}
\tablecolumns{4}
\tablecaption{Orion extinction curve presented as $f_{\lambda}$ for select emission line wavelengths \label{flambdatab}}
\tablehead{
\colhead{${\lambda}$ (\AA)} & \colhead{$x$ ($\mu$m$^{-1}$)} &
\colhead{ID} & \colhead{$f_{\lambda}$}
}
\startdata
1909 & 5.238 & {\rm C~{\sc iii}]} & 0.299 \\
2470 & 4.049 & {\rm [O~{\sc ii}]} & 0.288 \\
2697 & 3.708 & {\rm He~{\sc i}} & 0.226 \\
2723 & 3.672 & {\rm He~{\sc i}} & 0.222 \\
2764 & 3.618 & {\rm He~{\sc i}} & 0.217 \\
2829 & 3.535 & {\rm He~{\sc i}} & 0.209 \\
2945 & 3.396 & {\rm He~{\sc i}} & 0.199 \\
3188 & 3.137 & {\rm He~{\sc i}} & 0.195 \\
3697 & 2.705 & {\rm H~{\sc i}} & 0.163 \\
3704 & 2.700 & {\rm H~{\sc i}} & 0.162 \\
3712 & 2.694 & {\rm H~{\sc i}} & 0.161 \\
3722 & 2.687 & {\rm H~{\sc i}} & 0.160 \\
3729 & 2.682 & {\rm [O~{\sc ii}]} & 0.160 \\
3734 & 2.678 & {\rm H~{\sc i}} & 0.159 \\
3750 & 2.667 & {\rm H~{\sc i}} & 0.158 \\
3771 & 2.652 & {\rm H~{\sc i}} & 0.155 \\
3798 & 2.633 & {\rm H~{\sc i}} & 0.152 \\
3820 & 2.618 & {\rm He~{\sc i}} & 0.150 \\
3835 & 2.608 & {\rm H~{\sc i}} & 0.148 \\
3889 & 2.571 & {\rm He~{\sc i}} & 0.142 \\
3965 & 2.522 & {\rm He~{\sc i}} & 0.134 \\
3970 & 2.519 & {\rm H~{\sc i}} & 0.133 \\
4026 & 2.484 & {\rm He~{\sc i}} & 0.126 \\
4102 & 2.438 & {\rm H~{\sc i}} & 0.117 \\
4340 & 2.304 & {\rm H~{\sc i}} & 0.086 \\
4363 & 2.292 & {\rm [O~{\sc iii}]} & 0.082 \\
4388 & 2.279 & {\rm He~{\sc i}} & 0.078 \\
4400 & 2.273 & $B band$ & 0.076 \\
4471 & 2.237 & {\rm He~{\sc i}} & 0.064 \\
4713 & 2.122 & {\rm He~{\sc i}} & 0.024 \\
4861 & 2.057 & {\rm H~{\sc i}} & 0.000 \\
4922 & 2.032 & {\rm He~{\sc i}} & -0.009 \\
4959 & 2.017 & {\rm [O~{\sc iii}]} & -0.015 \\
5007 & 1.997 & {\rm [O~{\sc iii}]} & -0.022 \\
5016 & 1.994 & {\rm He~{\sc i}} & -0.024 \\
5500 & 1.818 & $V band$ & -0.091 \\
5518 & 1.812 & {\rm [Cl~{\sc iii}]} & -0.093 \\
5538 & 1.806 & {\rm [Cl~{\sc iii}]} & -0.096 \\
5755 & 1.738 & {\rm [N~{\sc ii}]} & -0.123 \\
5876 & 1.702 & {\rm He~{\sc i}} & -0.138 \\
6548 & 1.527 & {\rm [N~{\sc ii}]} & -0.218 \\
6563 & 1.524 & {\rm H~{\sc i}} & -0.220 \\
6583 & 1.519 & {\rm [N~{\sc ii}]} & -0.222 \\
6678 & 1.497 & {\rm He~{\sc i}} & -0.233 \\
6716 & 1.489 & {\rm [S~{\sc ii}]} & -0.238 \\
6731 & 1.486 & {\rm [S~{\sc ii}]} & -0.239 \\
7065 & 1.415 & {\rm He~{\sc i}} & -0.278 \\
7136 & 1.401 & {\rm [Ar~{\sc iii}]} & -0.286 \\
7236 & 1.382 & {\rm C~{\sc ii}} & -0.298 \\
7254 & 1.379 & {\rm O~{\sc i}} & -0.300 \\
7281 & 1.373 & {\rm He~{\sc i}} & -0.303 \\
7319 & 1.366 & {\rm [O~{\sc ii}]} & -0.307 \\
7330 & 1.364 & {\rm [O~{\sc ii}]} & -0.309 \\
9229 & 1.084 & {\rm H~{\sc i}} & -0.507 \\
9464 & 1.057 & {\rm He~{\sc i}} & -0.529 \\
9546 & 1.048 & {\rm H~{\sc i}} & -0.537 \\
9603 & 1.041 & {\rm He~{\sc i}} & -0.542 \\
10049 & 0.995 & {\rm H~{\sc i}} & -0.579 \\
10311 & 0.970 & {\rm He~{\sc i}} & -0.598
\enddata
\end{deluxetable}

%% file: tab3.tex
\begin{deluxetable}{lrrrc}
\tabletypesize{\normalsize}
\tablewidth{0pt}
\tablecolumns{5}
\tablecaption{Observed ratios of common-upper-level lines, $I_{2471}/I_{7325}$, and predictions from theory \label{2471to7325}}
\tablehead{
\colhead{Dataset} & \multicolumn{2}{c}{Theory} & 
\colhead{Uncorrected} & \colhead{Corrected for reddening} \\
\colhead{} & \colhead{Z82\tablenotemark{a}} & \colhead{W96\tablenotemark{b}} & \colhead{} & \colhead{using H~{\sc i} lines}
}
\startdata
FOS-1SW & 0.75 & 0.81  & $ 0.35\pm0.01$ & $ 0.71\pm0.01$\\
\\
STIS-SLIT1b & 0.75 & 0.81  & $ 0.19\pm0.01$ & $ 0.61\pm0.02$\\
STIS-SLIT1c & 0.75 & 0.81  & $ 0.27\pm0.01$ & $ 0.66\pm0.01$\enddata
\tablenotetext{a}{Using \citet{zei82} transition probabilities}
\tablenotetext{b}{Using \citet{wie96} transition probabilities}
\end{deluxetable}

%% file: tab4.tex
\begin{deluxetable}{rcrrrrrrrrr}
\tabletypesize{\tiny}
\setlength{\tabcolsep}{0.03in}
\tablewidth{0pt}
\tablecolumns{11}
\tablecaption{He~{\sc i} relative intensities, $I_{\lambda}/I_{4471}$, after correcting for reddening using H~{\sc i} Balmer (and Paschen) lines and our extinction curve \label{all_helium_table}}
\tablehead{
\colhead{$\lambda$} & \colhead{$n$} & \multicolumn{3}{c}{Predicted} & \multicolumn{6}{c}{Observed} \\
\colhead{(\AA)} & \colhead{} & \colhead{Case B} & \colhead{Model M} & \colhead{Model K} & \colhead{OTV92} & \colhead{EPTE98} & 
\colhead{BVV00} & \colhead{EPG04} & \colhead{FOS-1SW} & \colhead{STIS-SLIT1c} \\
\colhead{(1)} & \colhead{(2)} & \colhead{(3)} & \colhead{(4)} & \colhead{(5)} & \colhead{(6)} & 
\colhead{(7)} & \colhead{(8)} & \colhead{(9)} & \colhead{(10)} & \colhead{(11)}
}
\startdata
\multicolumn{11}{c}{2~$^3S~-~n~^3P$} \\*
\hline
3889 &  3 & 2.315 & 0.780 & 0.580 &  $1.019\pm0.362$ &  $1.165\pm0.242$ &  $1.126\pm0.191$ &  $0.950\pm0.152$ &  $1.005\pm0.131$ & \nodata \\+H~{\sc i} & \nodata &  4.575\tablenotemark{a}  & 3.389 & 2.651 & $3.112\pm0.561$ & $3.518\pm0.393$ & $3.702\pm0.163$ & $3.267\pm0.103$ & $3.425\pm0.080$ & \nodata \\*
3188 &  4 & 0.878 & 0.441 & 0.337 & $0.933\pm0.487$ & \nodata & \nodata & $0.536\pm0.043$ & $0.521\pm0.018$ & \nodata \\*
2945 &  5 & 0.414 & 0.290 & 0.229 & \nodata & \nodata & \nodata & \nodata & $0.310\pm0.017$ & $0.257\pm0.010$ \\*
2829 &  6 & 0.228 & 0.200 & 0.162 & \nodata & \nodata & \nodata & \nodata & \nodata & \nodata \\+[Fe~{\sc iv}] & \nodata & \nodata  & 0.268 & 0.248 & \nodata & \nodata & \nodata & \nodata & $0.232\pm0.017$ & \nodata \\*
2764 &  7 & 0.141 & 0.140 & 0.116 & \nodata & \nodata & \nodata & \nodata & $0.140\pm0.030$ & $0.120\pm0.004$ \\*
2723 &  8 & 0.093 & 0.100 & 0.085 & \nodata & \nodata & \nodata & \nodata & $0.093\pm0.028$ & $0.085\pm0.003$ \\*
2696 &  9 & 0.065 & 0.073 & 0.063 & \nodata & \nodata & \nodata & \nodata & $0.048\pm0.026$ & \nodata \\
\hline 
\multicolumn{11}{c}{2~$^1S~-~n~^1P$} \\*
\hline
5016 &  3 & 0.566 & 0.517 & 0.488 & \nodata & $0.496\pm0.070$ & $0.551\pm0.014$ & $0.502\pm0.007$ & \nodata & $0.473\pm0.013$ \\*
3965 &  4 & 0.221 & 0.206 & 0.193 & \nodata & $0.201\pm0.045$ & $0.236\pm0.006$ & $0.209\pm0.007$ & \nodata & \nodata \\+H~{\sc i}/[Ne~{\sc iii}] & \nodata & \nodata  & 4.973 & 4.187 & $4.405\pm0.794$ & \nodata & \nodata & \nodata & $4.924\pm0.117$ & \nodata \\*
3614 &  5 & 0.108 & 0.104 & 0.095 & $0.146\pm0.037$ & $0.085\pm0.019$ & $0.109\pm0.003$ & $0.104\pm0.007$ & \nodata & \nodata \\*
3448 &  6 & 0.059 & 0.061 & 0.052 & $0.105\pm0.055$ & \nodata & \nodata & $0.068\pm0.006$ & \nodata & \nodata \\*
3355 &  7 & 0.037 & 0.041 & 0.032 & $0.208\pm0.109$ & \nodata & \nodata & $0.042\pm0.006$ & \nodata & \nodata \\*
3297 &  8 & 0.024 & 0.029 & 0.021 & \nodata & \nodata & \nodata & $0.027\pm0.008$ & \nodata & \nodata \\
\hline 
\multicolumn{11}{c}{2~$^3P~-~n~^3S$} \\*
\hline
7065 &  3 & 0.612 & 1.714 & 1.833 & $1.414\pm0.255$ & \nodata & $1.951\pm0.079$ & $1.832\pm0.130$ & $1.617\pm0.065$ & $1.458\pm0.024$ \\*
4713 &  4 & 0.103 & 0.153 & 0.157 & $0.150\pm0.038$ & $0.145\pm0.032$ & $0.153\pm0.005$ & $0.151\pm0.002$ & \nodata & \nodata \\*
4121 &  5 & 0.038 & 0.046 & 0.046 & $0.046\pm0.013$ & $0.049\pm0.011$ & $0.044\pm0.001$ & $0.049\pm0.002$ & \nodata & \nodata \\*
3868 &  6 & 0.018 & 0.020 & 0.020 & \nodata & \nodata & $0.019\pm0.003$ & $0.017\pm0.002$ & \nodata & \nodata \\+[Ne~{\sc iii}] & \nodata & \nodata  & 2.846 & 2.846 & $3.382\pm0.610$ & \nodata & \nodata & \nodata & \nodata & \nodata \\*
3733 &  7 & 0.010 & 0.011 & 0.011 & \nodata & \nodata & $0.010\pm0.001$ & $0.011\pm0.004$ & \nodata & \nodata \\*
3652 &  8 & 0.006 & 0.007 & 0.007 & \nodata & \nodata & \nodata & $0.005\pm0.002$ & \nodata & \nodata \\*
3599 &  9 & 0.004 & 0.004 & 0.004 & \nodata & \nodata & \nodata & \nodata & \nodata & \nodata \\*
3563 & 10 & 0.003 & 0.003 & 0.003 & \nodata & \nodata & \nodata & \nodata & \nodata & \nodata \\*
3537 & 11 & 0.002 & 0.002 & 0.002 & \nodata & \nodata & \nodata & $0.003\pm0.001$ & \nodata & \nodata \\
\hline 
\multicolumn{11}{c}{2~$^3P~-~n~^3D$} \\*
\hline
5876 &  3 & 2.789 & 2.815 & 2.820 & $2.736\pm0.493$ & $3.119\pm0.349$ & $3.255\pm0.121$ & $3.165\pm0.100$ & $2.731\pm0.058$ & $2.902\pm0.044$ \\*
4471 &  4 & 1.000 & 1.000 & 1.000 & $1.000\pm0.150$ & $1.000\pm0.100$ & $1.000\pm0.023$ & $1.000\pm0.010$ & $1.000\pm0.019$ & $1.000\pm0.011$ \\*
4026 &  5 & 0.472 & 0.466 & 0.466 & $0.484\pm0.103$ & $0.440\pm0.062$ & $0.497\pm0.014$ & $0.479\pm0.015$ & $0.431\pm0.021$ & \nodata \\*
3820 &  6 & 0.254 & 0.254 & 0.254 & $0.237\pm0.059$ & $0.240\pm0.054$ & $0.232\pm0.006$ & $0.262\pm0.008$ & $0.227\pm0.027$ & \nodata \\*
3705 &  7 & 0.155 & 0.155 & 0.155 & Blend & \nodata & $0.176\pm0.005$ & $0.153\pm0.008$ & \nodata & \nodata \\*
3634 &  8 & 0.102 & 0.102 & 0.102 & $0.129\pm0.038$ & $0.096\pm0.021$ & $0.129\pm0.004$ & $0.104\pm0.007$ & \nodata & \nodata \\*
3587 &  9 & 0.071 & 0.071 & 0.071 & $0.090\pm0.026$ & $0.075\pm0.017$ & $0.083\pm0.003$ & $0.071\pm0.006$ & \nodata & \nodata \\*
3554 & 10 & 0.052 & 0.051 & 0.051 & $0.071\pm0.021$ & $0.038\pm0.008$ & $0.057\pm0.002$ & $0.050\pm0.005$ & \nodata & \nodata \\*
3531 & 11 & 0.038 & 0.038 & 0.037 & $0.060\pm0.017$ & \nodata & $0.037\pm0.002$ & $0.039\pm0.007$ & \nodata & \nodata \\*
3513 & 12 & 0.029 & 0.029 & 0.029 & $0.046\pm0.013$ & \nodata & $0.034\pm0.002$ & $0.028\pm0.005$ & \nodata & \nodata \\*
3499 & 13 & 0.023 & 0.023 & 0.023 & $0.042\pm0.022$ & \nodata & $0.030\pm0.002$ & $0.023\pm0.005$ & \nodata & \nodata \\*
3488 & 14 & 0.019 & 0.018 & 0.018 & $0.028\pm0.014$ & \nodata & \nodata & $0.018\pm0.004$ & \nodata & \nodata \\*
3479 & 15 & 0.015 & 0.015 & 0.015 & $0.015\pm0.008$ & \nodata & \nodata & $0.013\pm0.003$ & \nodata & \nodata \\*
3474 & 16 & 0.012 & 0.013 & 0.012 & \nodata & \nodata & \nodata & $0.013\pm0.004$ & \nodata & \nodata \\*
3468 & 17 & 0.010 & 0.011 & 0.011 & \nodata & \nodata & \nodata & $0.007\pm0.003$ & \nodata & \nodata \\
\hline 
\multicolumn{11}{c}{2~$^1P~-~n~^1S$} \\*
\hline
7281 &  3 & 0.151 & 0.155 & 0.145 & $0.129\pm0.032$ & \nodata & $0.153\pm0.004$ & $0.152\pm0.012$ & \nodata & $0.148\pm0.002$ \\*
5048 &  4 & 0.038 & 0.040 & 0.036 & \nodata & $0.042\pm0.009$ & $0.043\pm0.002$ & $0.126\pm0.003$ & \nodata & \nodata \\*
4438 &  5 & 0.015 & 0.016 & 0.015 & $0.017\pm0.006$ & $0.013\pm0.004$ & $0.0149\pm0.0004$ & $0.016\pm0.001$ & \nodata & \nodata \\*
4169 &  6 & 0.008 & 0.008 & 0.007 & $0.009\pm0.004$ & $0.011\pm0.003$ & $0.0111\pm0.0008$ & $0.013\pm0.001$ & \nodata & \nodata \\*
4024 &  7 & 0.005 & 0.005 & 0.004 & \nodata & \nodata & $0.0055\pm0.0005$ & \nodata & \nodata & \nodata \\
\hline 
\tablebreak
\multicolumn{11}{c}{2~$^1P~-~n~^1D$} \\*
\hline
6678 &  3 & 0.784 & 0.737 & 0.731 & $0.704\pm0.149$ & $0.939\pm0.133$ & $0.955\pm0.024$ & $0.912\pm0.055$ & $0.729\pm0.031$ & $0.743\pm0.013$ \\*
4922 &  4 & 0.272 & 0.258 & 0.256 & $0.238\pm0.051$ & $0.263\pm0.037$ & $0.289\pm0.007$ & $0.267\pm0.004$ & $0.265\pm0.028$ & $0.204\pm0.005$ \\*
4388 &  5 & 0.125 & 0.120 & 0.119 & $0.112\pm0.033$ & $0.117\pm0.026$ & $0.121\pm0.003$ & $0.120\pm0.003$ & $0.103\pm0.016$ & $0.105\pm0.003$ \\*
4144 &  6 & 0.067 & 0.066 & 0.065 & $0.058\pm0.017$ & $0.060\pm0.013$ & $0.067\pm0.002$ & $0.063\pm0.003$ & \nodata & \nodata \\*
4009 &  7 & 0.040 & 0.040 & 0.040 & $0.042\pm0.012$ & $0.046\pm0.010$ & $0.040\pm0.001$ & $0.037\pm0.002$ & \nodata & \nodata \\*
3927 &  8 & 0.026 & 0.026 & 0.026 & \nodata & \nodata & \nodata & \nodata & \nodata & \nodata \\
\hline 
\multicolumn{11}{c}{3~$^3S~-~n~^3P$} \\*
\hline
9464 &  5 & 0.023 & 0.053 & 0.054 & $0.080\pm0.020$ & \nodata & \nodata & $0.027\pm0.004$ & \nodata & \nodata \\*
8362 &  6 & 0.015 & 0.030 & 0.029 & Blend & \nodata & \nodata & $0.033\pm0.004$ & \nodata & \nodata \\*
7816 &  7 & 0.010 & 0.018 & 0.017 & $0.014\pm0.004$ & \nodata & \nodata & $0.022\pm0.002$ & \nodata & \nodata \\*
7500 &  8 & 0.007 & 0.012 & 0.011 & $0.012\pm0.004$ & \nodata & \nodata & $0.014\pm0.001$ & \nodata & \nodata \\*
7298 &  9 & 0.005 & 0.008 & 0.007 & $0.008\pm0.003$ & \nodata & $0.0090\pm0.0003$ & $0.0095\pm0.0010$ & \nodata & \nodata \\*
7161 & 10 & 0.004 & 0.006 & 0.005 & \nodata & \nodata & $0.0067\pm0.0002$ & $0.0070\pm0.0007$ & \nodata & \nodata \\*
7062 & 11 & 0.003 & 0.004 & 0.004 & \nodata & \nodata & $0.0046\pm0.0002$ & $0.0048\pm0.0005$ & \nodata & \nodata \\*
6989 & 12 & 0.002 & 0.003 & 0.003 & \nodata & \nodata & $0.0030\pm0.0001$ & $0.0032\pm0.0004$ & \nodata & \nodata \\*
6934 & 13 & 0.002 & 0.003 & 0.002 & \nodata & \nodata & $0.0025\pm0.0002$ & $0.0033\pm0.0005$ & \nodata & \nodata \\*
6890 & 14 & 0.001 & 0.002 & 0.002 & \nodata & \nodata & \nodata & \nodata & \nodata & \nodata \\*
6856 & 15 & 0.001 & 0.002 & 0.002 & \nodata & \nodata & \nodata & $0.0022\pm0.0004$ & \nodata & \nodata \\
\hline 
\multicolumn{11}{c}{3~$^1S~-~n~^1P$} \\*
\hline
9603 &  6 & 0.005 & 0.006 & 0.005 & $0.011\pm0.003$ & \nodata & \nodata & \nodata & \nodata & \nodata \\*
8915 &  7 & 0.004 & 0.004 & 0.003 & $0.007\pm0.002$ & \nodata & \nodata & $0.0057\pm0.0009$ & \nodata & \nodata \\*
8518 &  8 & 0.003 & 0.003 & 0.002 & $0.004\pm0.002$ & \nodata & \nodata & $0.0029\pm0.0005$ & \nodata & \nodata \\*
8266 &  9 & 0.002 & 0.002 & 0.002 & Blend & \nodata & \nodata & \nodata & \nodata & \nodata \\*
8094 & 10 & 0.001 & 0.002 & 0.001 & $0.005\pm0.002$ & \nodata & \nodata & $0.0015\pm0.0003$ & \nodata & \nodata \\*
7972 & 11 & 0.001 & 0.002 & \nodata & \nodata & \nodata & \nodata & $0.0012\pm0.0003$ & \nodata & \nodata \\
\hline 
\multicolumn{11}{c}{3~$^3P~-~n~^3D$} \\*
\hline
10311 &  6 & 0.029 & 0.029 & 0.029 & $0.012\pm0.006$ & \nodata & \nodata & $0.038\pm0.006$ & \nodata & \nodata \\*
9517 &  7 & 0.019 & 0.019 & 0.019 & \nodata & \nodata & \nodata & $0.009\pm0.001$ & \nodata & \nodata \\*
9063 &  8 & 0.013 & 0.013 & 0.013 & \nodata & \nodata & \nodata & $0.015\pm0.002$ & \nodata & \nodata \\*
8777 &  9 & 0.009 & 0.009 & 0.009 & $0.010\pm0.003$ & \nodata & \nodata & $0.024\pm0.003$ & \nodata & \nodata \\*
8583 & 10 & 0.007 & 0.007 & 0.007 & \nodata & \nodata & \nodata & \nodata & \nodata & \nodata \\*
8445 & 11 & 0.005 & 0.005 & 0.005 & \nodata & \nodata & \nodata & \nodata & \nodata & \nodata \\*
8342 & 12 & 0.004 & 0.004 & 0.004 & \nodata & \nodata & \nodata & $0.0067\pm0.0009$ & \nodata & \nodata \\*
8265 & 13 & 0.003 & 0.003 & 0.003 & Blend & \nodata & \nodata & \nodata & \nodata & \nodata \\*
8204 & 14 & 0.003 & 0.003 & 0.003 & \nodata & \nodata & \nodata & $0.0026\pm0.0004$ & \nodata & \nodata \\*
8156 & 15 & 0.002 & 0.002 & 0.002 & \nodata & \nodata & \nodata & $0.0022\pm0.0004$ & \nodata & \nodata \\*
8116 & 16 & 0.002 & 0.002 & 0.002 & \nodata & \nodata & \nodata & $0.0016\pm0.0003$ & \nodata & \nodata \\*
8084 & 17 & 0.001 & 0.002 & 0.002 & \nodata & \nodata & \nodata & $0.0007\pm0.0003$ & \nodata & \nodata \\
\hline 
\multicolumn{5}{l}{$C_{\mathrm{H}\beta}$ (from H~{\sc i} lines)}  & $ 0.61\pm0.11$ & $ 0.50\pm0.21$ & $ 0.30\pm0.12$ & $ 0.82\pm0.04$ & $ 0.52\pm0.05$ & $ 0.66\pm0.04$ \\
\\
\multicolumn{5}{l}{$C_{\mathrm{H}\beta}$ (from H~{\sc i} + He~{\sc i} subset)}  & $ 0.57\pm0.06$ & $ 0.60\pm0.12$ & $ 0.31\pm0.07$ & $ 0.82\pm0.07$ & $ 0.52\pm0.03$ & $ 0.69\pm0.05$ \\
\multicolumn{5}{l}{He$^+$/H$^+$ (from H~{\sc i} + He~{\sc i} subset)}   & $0.098\pm0.007$ & $0.084\pm0.007$ & $0.081\pm0.005$ & $0.088\pm0.003$ & $0.083\pm0.002$ & $0.095\pm0.003$ \\
\\
\multicolumn{5}{l}{He$^+$/H$^+$ (from source, select He~{\sc i} lines)}   & $0.089\pm0.002$ & $0.089\pm0.009$ & \nodata & $0.087\pm0.001$ & \nodata & \nodata
\enddata
\tablenotetext{a}{Case~B H~{\sc i} + He~{\sc i} blend determined using a typical He$^{+}$/H$^{+}$ (0.088) as found from these data}
\end{deluxetable}

%% file: tab5.tex
\begin{deluxetable}{lrr}
\tablewidth{0pt}
\tablecolumns{3}
\tablecaption{CLOUDY input parameters\label{cloudy_hei}}
\tablehead{   
\colhead{Quantity} &
\colhead{Model M\tablenotemark{a}} & \colhead{Model K\tablenotemark{b}}
}
\startdata
$T_{\mathrm{eff}}$ (K) & 35200 & 41200 \\
log$\phi$(H) &  13.05 &  13.10 \\
radius (pc) &   0.27 &   0.27 \\
log$n_{\mathrm H}\mathrm{(inner)}$ (cm$^{-3}$) &  3.4 &  3.4\\
Turbulence (km~s$^{-1}$) & 12 & 12 \\
\enddata
\tablenotetext{a}{\citet{mih72} stellar atmosphere}
\tablenotetext{b}{\citet{kur79} stellar atmosphere}
\end{deluxetable}

%% file: tab6.tex
\begin{deluxetable}{lrr}
\tabletypesize{\small}
\tablewidth{0pt}
\tablecolumns{3}
\tablecaption{Model abundances relative to H (12+log($X$/H))\label{abund_table_hei}}
\tablehead{
\colhead{Element} &
\colhead{Model M\tablenotemark{a}} & \colhead{Model K\tablenotemark{b}} \\
\colhead{(1)} & \colhead{(2)} & \colhead{(3)}
}
\startdata
He &      10.98 &      10.98 \\
C  &       8.48 &       8.42 \\
N  &       7.85 &       7.73 \\
O  &       8.60 &       8.65 \\
Ne &       7.78 &       8.05 \\
S  &       7.00 &       7.22 \\
Ar &       6.48 &       6.62 \\
Cl &       5.00 &       5.46 \\
Fe &       6.48 &       6.48 \\
\enddata
\tablenotetext{a}{CLOUDY H~{\sc ii} region abundances from \citet{bal91,rub91} and OTV92}
\tablenotetext{b}{EPG04 abundances}
\end{deluxetable}

%% file: tab7.tex
\begin{deluxetable}{lcrr}
\tabletypesize{\small}
\tablewidth{0pt}
\tablecolumns{4}
\tablecaption{Constraints on Model Parameters \label{constraint_hei}}
\tablehead{
\colhead{Quantity} & 
\multicolumn{3}{c}{Orion Nebula (1SW)} \\
\cline{2-4} \\
\colhead{} & \colhead{SLIT1c} & 
\colhead{Model M} & \colhead{Model K} 
}
\startdata
 SB(H$\beta$)\tablenotemark{a} & $    66.8\pm0.7$ &     64.1 &     63.2 \\
 $\lambda5007$/H$\beta$ & $    3.13\pm0.05$ &      3.5 &      4.3 \\
 $\lambda5007/\lambda3726$\tablenotemark{b} & $     4.0\pm0.1$ &      5.9 &      6.6 \\
 $(\lambda4959+\lambda5007)/\lambda4363$\tablenotemark{c} & $     389\pm7$ &    421.0 &    392.1 \\
 $(\lambda6548+\lambda6583)/\lambda5755$\tablenotemark{c} & $      64\pm1$ &     61.6 &     65.0 \\
 $\lambda6731/\lambda6716$\tablenotemark{d} & $    2.00\pm0.04$ &      2.1 &      2.0 \\
 $\lambda3726/\lambda3729$\tablenotemark{d} & $     2.3\pm0.2$ &      2.5 &      2.4 \\
 $\lambda1907/\lambda1909$\tablenotemark{d} & $    1.20\pm0.04$ &      1.1 &      1.2 \\
 $(\lambda3726+\lambda3729)/\lambda7325$\tablenotemark{d} & $     6.4\pm0.6$ &      4.7 &      6.0 \\
\enddata
\tablenotetext{a}{Surface brightness in units of $10^{-13}$ ergs cm$^{-2}$ s$^{-1}$ arcsec$^{-2}$}
\tablenotetext{b}{Ionization indicator}
\tablenotetext{c}{Electron temperature indicator}
\tablenotetext{d}{Electron density indicator}
\end{deluxetable}

%% file: tab8.tex
\begin{deluxetable}{clrrrrr}
\tabletypesize{\tiny}
\tablewidth{0pt}
\tablecolumns{7}
\tablecaption{Dereddened HST STIS V2 and FOS observations ($I_{\lambda}/I_{\mathrm{H}\beta, predicted}$) \label{corrected_table}}
\tablehead{
\colhead{ID} & \colhead{$\lambda$ (\AA)} & 
\colhead{SLIT1b} & \colhead{SLIT1c} & 
\colhead{SLIT2b} & \colhead{SLIT2c} & \colhead{FOS-1SW}
}
\startdata
{\rm C~{\sc iii}]} & $1907$ & $  0.0691\pm0.0038$ & $  0.0872\pm0.0029$ & \nodata & \nodata & \nodata \\
{\rm C~{\sc iii}]} & $1909$ & $  0.0591\pm0.0038$ & $  0.0726\pm0.0022$ & \nodata & \nodata & \nodata \\
{\rm [O~{\sc ii}]} & $2470$\tablenotemark{c} & $  0.0718\pm0.0023$ & $  0.1208\pm0.0035$ & $  0.0550\pm0.0024$ & $  0.0807\pm0.0032$ & $  0.1613\pm0.0024$ \\
{\rm He~{\sc i}} & $2697$\tablenotemark{c} & \nodata & \nodata & \nodata & \nodata & $0.0021\pm0.0012  $ \\
{\rm He~{\sc i}} & $2723$\tablenotemark{c} & $  0.0036\pm0.0078$ & $  0.0042\pm0.0002$ & \nodata & \nodata & $  0.0041\pm0.0012$ \\
{\rm He~{\sc i}} & $2764$\tablenotemark{c} & $  0.0075\pm0.0007$ & $  0.0059\pm0.0002$ & \nodata & \nodata & $  0.0061\pm0.0013$ \\
{\rm He~{\sc i}}+{\rm [Fe~{\sc iv}]} & $2829$ & \nodata & \nodata & \nodata & \nodata & $0.0102\pm0.0007$ \\
{\rm He~{\sc i}} & $2945$ & $  0.0126\pm0.0011$ & $  0.0126\pm0.0006$ & \nodata & \nodata & $  0.0136\pm0.0007$ \\
{\rm He~{\sc i}} & $3188$ & \nodata & \nodata & \nodata & \nodata & $0.0229\pm0.0007$ \\
{\rm H~{\sc i}} & $3697$\tablenotemark{c} & $  0.0161\pm0.0326$ & $  0.0089\pm0.0006$ & $  0.0175\pm0.0140$ & $  0.0061\pm0.0123$ & \nodata \\
{\rm H~{\sc i}} & $3704$\tablenotemark{c} & $  0.0152\pm0.0137$ & $  0.0187\pm0.0080$ & $  0.0256\pm0.0057$ & $  0.0194\pm0.0025$ & \nodata \\
{\rm H~{\sc i}} & $3712$\tablenotemark{c} & $  0.0153\pm0.0017$ & $  0.0182\pm0.0009$ & $  0.0138\pm0.0171$ & $  0.0139\pm0.0042$ & \nodata \\
{\rm H~{\sc i}}+{\rm [S~{\sc iii}]} & $3722$ & $  0.0298\pm0.0015$ & $  0.0363\pm0.0011$ & $  0.0305\pm0.0010$ & $  0.0370\pm0.0025$ & \nodata \\
{\rm [O~{\sc ii}]} & $3726$ & $  0.6725\pm0.0199$ & $  0.7875\pm0.0267$ & $  0.6690\pm0.0187$ & $  0.7199\pm0.0330$ & \nodata \\
{\rm [O~{\sc ii}]} & $3729$ & $  0.3234\pm0.0298$ & $  0.3451\pm0.0298$ & $  0.3353\pm0.0252$ & $  0.3503\pm0.0199$ & \nodata \\
{\rm H~{\sc i}} & $3734$\tablenotemark{c} & $  0.0297\pm0.0027$ & $  0.0266\pm0.0013$ & $  0.0201\pm0.0007$ & $  0.0247\pm0.0018$ & \nodata \\
{\rm H~{\sc i}} & $3750$\tablenotemark{c} & $  0.0265\pm0.0022$ & $  0.0306\pm0.0011$ & $  0.0330\pm0.0028$ & $  0.0260\pm0.0013$ & $  0.0285\pm0.0018$ \\
{\rm H~{\sc i}} & $3771$\tablenotemark{c} & $  0.0319\pm0.0015$ & $  0.0387\pm0.0011$ & $  0.0380\pm0.0014$ & $  0.0377\pm0.0015$ & $  0.0397\pm0.0014$ \\
{\rm H~{\sc i}} & $3798$\tablenotemark{c} & $  0.0425\pm0.0011$ & $  0.0509\pm0.0014$ & $  0.0400\pm0.0012$ & $  0.0490\pm0.0019$ & $  0.0520\pm0.0015$ \\
{\rm He~{\sc i}} & $3820$\tablenotemark{c} & \nodata & \nodata & \nodata & \nodata & $ 0.0099\pm0.0012  $ \\
{\rm He~{\sc i}}+{\rm H~{\sc i}} & $3889$ & \nodata & \nodata & \nodata & \nodata & $0.1499\pm0.0027  $ \\
{\rm He~{\sc i}}+{\rm H~{\sc i}}+{\rm [Ne~{\sc iii}]} & $3965$ & \nodata & \nodata & \nodata & \nodata & $0.2155\pm0.0041  $ \\
{\rm He~{\sc i}} & $4026$\tablenotemark{c} & \nodata & \nodata & \nodata & \nodata & $0.0188\pm0.0009  $ \\
{\rm H~{\sc i}} & $4340$\tablenotemark{c} & $  0.4817\pm0.0058$ & $  0.4817\pm0.0126$ & $  0.4542\pm0.0042$ & $  0.4822\pm0.0158$ & $  0.4784\pm0.0065$ \\
{\rm [O~{\sc iii}]} & $4363$ & $  0.0109\pm0.0003$ & $  0.0109\pm0.0003$ & $  0.0088\pm0.0003$ & $  0.0136\pm0.0005$ & \nodata \\
{\rm He~{\sc i}} & $4388$\tablenotemark{c} & $  0.0057\pm0.0002$ & $  0.0051\pm0.0002$ & $  0.0046\pm0.0002$ & $  0.0057\pm0.0002$ & $  0.0045\pm0.0007$ \\
{\rm He~{\sc i}} & $4471$\tablenotemark{c} & $  0.0457\pm0.0006$ & $  0.0481\pm0.0013$ & $  0.0415\pm0.0005$ & $  0.0452\pm0.0015$ & $  0.0436\pm0.0010$ \\
{\rm H~{\sc i}} & $4861$\tablenotemark{c} & $  1.0085\pm0.0112$ & $  1.0044\pm0.0260$ & $  1.0382\pm0.0076$ & $  0.9990\pm0.0323$ & $  0.9671\pm0.0146$ \\
{\rm He~{\sc i}} & $4922$\tablenotemark{c} & $  0.0092\pm0.0010$ & $  0.0097\pm0.0003$ & $  0.0145\pm0.0010$ & $  0.0104\pm0.0007$ & $  0.0115\pm0.0012$ \\
{\rm [O~{\sc iii}]} & $4959$ & $  1.2109\pm0.0145$ & $  1.0535\pm0.0272$ & $  1.2355\pm0.0097$ & $  1.1675\pm0.0380$ & \nodata \\
{\rm [O~{\sc iii}]} & $5007$ & $  3.6142\pm0.0393$ & $  3.1408\pm0.0809$ & $  3.6795\pm0.0269$ & $  3.4824\pm0.1124$ & \nodata \\
{\rm He~{\sc i}} & $5016$ & $  0.0260\pm0.0007$ & $  0.0226\pm0.0008$ & $  0.0213\pm0.0006$ & $  0.0196\pm0.0007$ & \nodata \\
{\rm [Cl~{\sc iii}]} & $5518$ & $  0.0040\pm0.0002$ & $  0.0033\pm0.0001$ & $  0.0025\pm0.0002$ & $  0.0030\pm0.0002$ & \nodata \\
{\rm [Cl~{\sc iii}]} & $5538$ & $  0.0062\pm0.0002$ & $  0.0055\pm0.0001$ & $  0.0043\pm0.0001$ & $  0.0052\pm0.0002$ & \nodata \\
{\rm [N~{\sc ii}]} & $5755$ & $  0.0074\pm0.0002$ & $  0.0122\pm0.0003$ & $  0.0068\pm0.0002$ & $  0.0084\pm0.0003$ & \nodata \\
{\rm He~{\sc i}} & $5876$\tablenotemark{c} & $  0.1363\pm0.0015$ & $  0.1365\pm0.0035$ & $  0.1345\pm0.0008$ & $  0.1371\pm0.0044$ & $  0.1179\pm0.0019$ \\
{\rm [N~{\sc ii}]} & $6548$ & $  0.1507\pm0.0020$ & $  0.1877\pm0.0049$ & $  0.1501\pm0.0015$ & $  0.1587\pm0.0052$ & \nodata \\
{\rm H~{\sc i}} & $6563$\tablenotemark{c} & $  2.8274\pm0.0327$ & $  2.7863\pm0.0741$ & $  2.8477\pm0.0209$ & $  2.8694\pm0.0949$ & $  3.0933\pm0.0466$\tablenotemark{a} \\
{\rm H~{\sc i}} & $6563$\tablenotemark{c} & \nodata & \nodata & \nodata & \nodata & $2.7514\pm0.0402$\tablenotemark{b} \\
{\rm [N~{\sc ii}]} & $6583$ & $  0.4601\pm0.0065$ & $  0.5834\pm0.0153$ & $  0.4582\pm0.0050$ & $  0.4896\pm0.0160$ & \nodata \\
{\rm He~{\sc i}} & $6678$\tablenotemark{c} & $  0.0343\pm0.0005$ & $  0.0346\pm0.0009$ & $  0.0332\pm0.0003$ & $  0.0337\pm0.0012$ & $  0.0317\pm0.0023$\tablenotemark{a} \\
{\rm He~{\sc i}} & $6678$\tablenotemark{c} & \nodata & \nodata & \nodata & \nodata & $0.0313\pm0.0012$\tablenotemark{b} \\
{\rm [S~{\sc ii}]} & $6716$ & $  0.0213\pm0.0005$ & $  0.0217\pm0.0006$ & $  0.0247\pm0.0005$ & $  0.0209\pm0.0007$ & \nodata \\
{\rm [S~{\sc ii}]} & $6731$ & $  0.0392\pm0.0007$ & $  0.0432\pm0.0012$ & $  0.0455\pm0.0006$ & $  0.0409\pm0.0014$ & \nodata \\
{\rm He~{\sc i}} & $7065$ & $  0.0669\pm0.0008$ & $  0.0675\pm0.0018$ & $  0.0668\pm0.0005$ & $  0.0668\pm0.0022$ & $  0.0694\pm0.0026$ \\
{\rm [Ar~{\sc iii}]} & $7136$ & $  0.1499\pm0.0019$ & $  0.1547\pm0.0041$ & $  0.1460\pm0.0013$ & $  0.1542\pm0.0051$ & \nodata \\
{\rm C~{\sc ii}} & $7236$ & $  0.0026\pm0.0002$ & $  0.0026\pm0.0002$ & $  0.0024\pm0.0001$ & $  0.0023\pm0.0002$ & \nodata \\
{\rm O~{\sc i}} & $7254$ & $  0.0012\pm0.0001$ & $  0.0011\pm0.0000$ & $  0.0009\pm0.0001$ & $  0.0012\pm0.0001$ & \nodata \\
{\rm He~{\sc i}} & $7281$\tablenotemark{c} & $  0.0065\pm0.0001$ & $  0.0068\pm0.0002$ & $  0.0058\pm0.0001$ & $  0.0061\pm0.0002$ & \nodata \\
{\rm [O~{\sc ii}]} & $7319$\tablenotemark{c} & $  0.0613\pm0.0008$ & $  0.0942\pm0.0025$ & $  0.0549\pm0.0006$ & $  0.0655\pm0.0022$ & $  0.2203\pm0.0047$ \\
{\rm [O~{\sc ii}]} & $7330$\tablenotemark{c} & $  0.0500\pm0.0006$ & $  0.0775\pm0.0020$ & $  0.0442\pm0.0005$ & $  0.0541\pm0.0018$ & Blend \\
\hline
 \multicolumn{2}{l}{$C_{\mathrm{H}\beta}$} & $0.89\pm0.05$ & $0.71\pm0.04$ & $0.74\pm0.08$ & $0.87\pm0.06$ & $0.54\pm0.02$ \\
 \multicolumn{2}{l}{He$^{+}$/H$^{+}$} & $  0.090\pm0.003$ & $  0.094\pm0.003$ & $  0.089\pm0.004$ & $  0.089\pm0.003$ & $  0.083\pm0.002$
\enddata
\tablenotetext{a}{FOS/G780 grating}
\tablenotetext{b}{FOS/G570 grating}
\tablenotetext{c}{Lines used to determine reddening curve.}
\end{deluxetable}